%% file: surf.tex
\documentclass[nonacm, sigconf,10pt]{acmart} 

\newcommand{\figref}[1]{Figure~\ref{#1}}
\newcommand{\secref}[1]{Section~\ref{#1}}
\newcommand{\tabref}[1]{Table~\ref{#1}}
\newcommand{\heading}[1]{\smallskip\noindent\textbf{#1}}

\usepackage{multirow}
\usepackage{epstopdf}

\usepackage{url} 

\usepackage[utf8]{inputenc}

\usepackage{array}

\usepackage{times}
\usepackage[english]{babel}
\usepackage{blindtext}
\usepackage{endnotes}
\usepackage{color,hyphenat}
\usepackage{listings}
\usepackage{booktabs}
\usepackage{algorithm,algpseudocode}
\usepackage[labelformat=simple]{subcaption}
\usepackage{stfloats}

\renewcommand\footnotetextcopyrightpermission[1]{} 
\setcopyright{none}
\definecolor{dkgreen}{rgb}{0,0.6,0}
\definecolor{gray}{rgb}{0.5,0.5,0.5}
\definecolor{mauve}{rgb}{0.58,0,0.82}

\algnewcommand\And{\textbf{and }}
\algnewcommand\Or{\textbf{or }}

\newcommand{\name}{{RF-Mediator}}

\begin{document}


\title[RF-Mediator]{Cross-Media Wireless Made Easier: \\ Tuning Media Interfaces with Flexible Metasurfaces}


\author{Ruichun Ma}
\affiliation{\institution{Yale University}}

\author{Wenjun Hu}
\affiliation{\institution{Yale University}}

\newcommand{\todoEdit}[1]{\textcolor{blue}{#1}}
\newcommand{\minorEdit}[1]{\textcolor{pink}{#1}}
\newcommand{\majorEdit}[1]{\textcolor{magenta}{#1}}
\newcommand{\wenjunEdit}[1]{\textcolor{red}{#1}}


\input{abstract}

\maketitle

\setlength{\abovedisplayskip}{2pt}
\setlength{\belowdisplayskip}{2pt}

\input{intro}
\input{motivation}
\input{design}
\input{implementation}

\input{eval}

\input{related}
\input{conclusion}

\clearpage
\bibliographystyle{concise2}
\bibliography{biblio}

\end{document}

%% file: abstract.tex


\begin{abstract}
Emerging wireless IoT applications increasingly venture beyond over-the-air communication, such as deep-tissue networking for implantable sensors, air-water communication for ocean monitoring, and soil sensing. These applications face the fundamental challenge of significant power loss due to reflection at media interfaces. 


We present \name, a programmable metasurface system placed at media interfaces to virtually mask the presence of the physical boundary. It is designed as a single-layer metasurface comprising arrays of varactor-based elements. By tuning the bias voltage element-wise, the surface mediates between media on both sides dynamically and beamforms towards the endpoint to boost transmission through the interface, as if no media interface existed. 
The control algorithm determines the surface configuration by probing the search space efficiently.
We fabricate the surface on a thin, flexible substrate, and experiment with several cross-media setups. 
Extensive evaluation shows that \name{} provides a median power gain of 8~dB for air-to-tissue links and up to 30~dB for cross-media backscatter links.

\end{abstract}

%% file: intro.tex
\section{Introduction}
\label{s:intro}

As new wireless Internet of Things (IoT) applications emerge, increasingly the communication media are not limited to air only. Numerous applications involve crossing over media interfaces, especially air-tissue interfaces~\cite{yu_magnetoelectric_mobicom22,inNout-mobicom20, umedic-mobicom20, remix-sigcomm18,ivn-sigcomm18,inter-tech-backscatter}, air-water interfaces~\cite{tarf-sigcomm18, amphilight-nsdi2020}, and air-ground interfaces~\cite{wireless_in_mines, ground_radar_2008}.
Air-tissue networking is the key enabler for numerous \textit{in-vivo} applications, 
such as internal vital sign monitoring~\cite{burton_wireless_2020, traverso_physiologic_2015}, wireless gastrointestinal diagnosis~\cite{meron_wireless_2000}, on-demand drug delivery~\cite{lee_biological_2015}, and untethered neuro-stimulation therapy~\cite{sun_closed-loop_2014, schuster_new_2016}.
Air-water networking arises in the course of marine biological sensing, underwater IoT, surveying and monitoring submerged sites~\cite{underwater-heritage}.
Air-ground networking has significance ont only for underground operation~\cite{wireless_in_mines, wireless-in-tunnel}, but also for soil sensing applications~\cite{ding_soil_sensing,khan_estimating_soil_2022}.

\begin{figure}[t]
    \centering
    \includegraphics[width=0.85\linewidth]{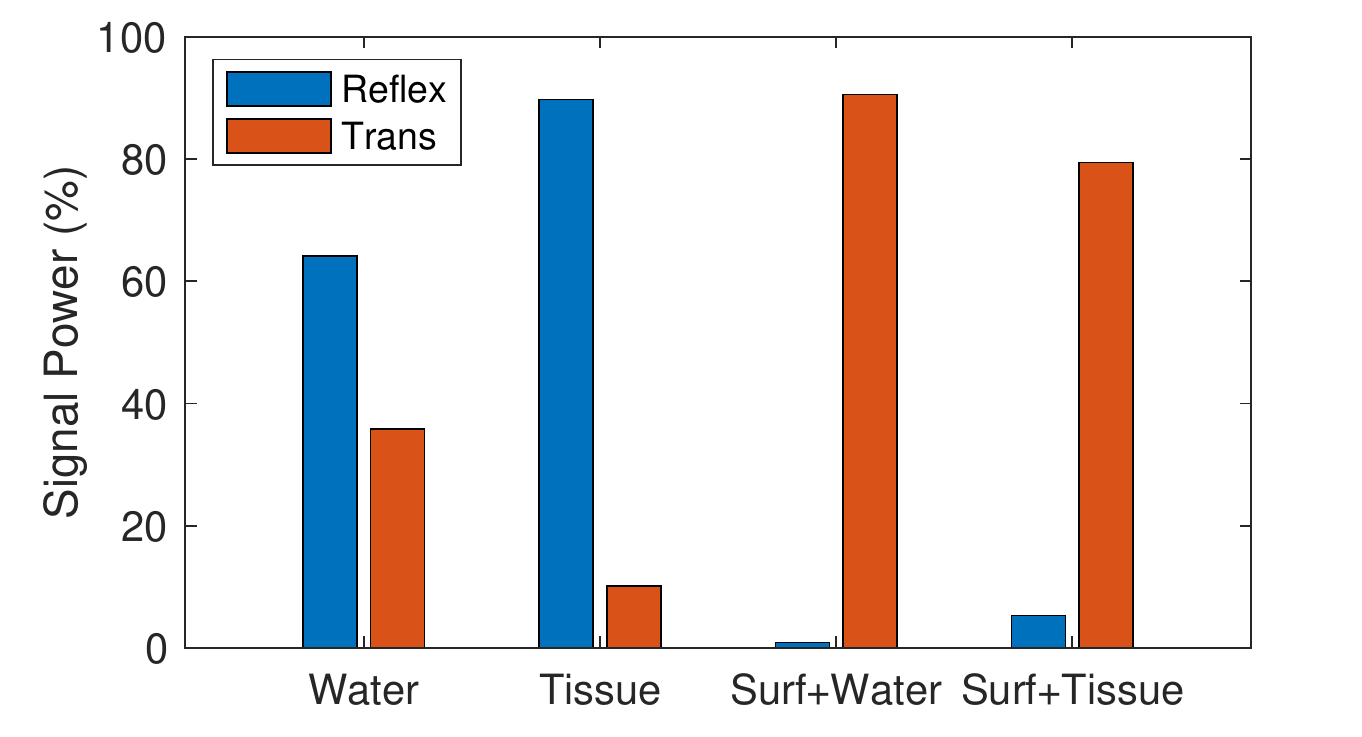}
    \caption{Reflected and Transmitted Signal Power w/ and w/o \name{}. \textmd{Our design can minimize reflection and boost through-interface transmission significantly as simulated with HFSS.} }
    \label{fig:motiv-trans-reflex-comparison}
\end{figure}

However, cross-media networking is much more challenging than its over-the-air counterpart (\secref{sec:motivation}). A fundamental challenge is that RF signals are susceptible to unfavorable propagation behavior when crossing media interfaces. When the two adjacent media exhibit different propagation characteristics,  
some amount of signal power is reflected at the interface rather than crossing the interface. In the case of air-tissue and air-water links, such reflection is substantial. On 2.4~GHz, around 90\% of the signal power is reflected (\figref{fig:motiv-trans-reflex-comparison}), amounting to 10~dB power reduction as the signals traverse the media interface.
This reflection happens in both propagation directions, which can incur notable power loss for in-vivo endpoints, especially for batteryless devices with low transmission power to start with. 
Strong reflections can further cause severe interference and potentially destructive multi-path fading for endpoints in the air.


Existing solutions tend to operate at the communication endpoints, enhancing the transmitted or received power via some form of beamforming~\cite{inNout-mobicom20, remix-sigcomm18, ivn-sigcomm18}. 
While these \textit{compensate for} the loss at the interface, they cannot address the root cause of the cross-media challenge, since the propagation behavior is determined by the physical properties of media. 


Fundamentally, different medium properties cause impedance mismatch at the interface, the root cause of the strong reflections~\cite{microwave-engineering} (\secref{sec:motiv:wave}). 
Consider media as transmission lines for signals, 
if we can place a component at the media interface for the matching purpose, e.g., extra antennas as the simplest components,
then the situation is conceptually similar to conventional impedance matching of RF circuit. 
The difference is that we perform impedance matching in the propagation environment to alter the unfavorable environment.

A smart surface is well suited to the role of impedance matching between media, the simplest of which is simply many antenna-like elements acting as a \textit{non-programmable surface}. 
There has been significant interest in augmenting or programming the radio propagation environment (e.g., ~\cite{press-hotnets17, laia-nsdi19, llama, rfocus, scattermimo, lava-sigcomm21}) by actuating passing wireless signals to control the end-to-end signal propagation behavior and improve the perceived received signals. However, these are designed for over-the-air propagation and cannot specifically alter the unfavorable propagation behavior when crossing a media interface. Further, their beamforming capability may degrade due to coupling with close-by non-air medium.
Recent work have made head way designing multi-layer metasurfaces for reflection mitigation~\cite{yang_near-reflectionless_2022, antireflection_cascaded}. However, these are one-off hardware designs far from handling real-world dynamics and lack consideration for wireless channels. Further, these designs forego the beamforming gain from previous endpoint-centric solutions. Practically, thick, hard circuit boards are ill-suited for deployment.

Instead, we present \name{}, 
a one-layer, programmable metasurface to be placed at the media interface (\secref{sec:design}). 
The surface hardware comprises metallic patterns interconnected with varactor diodes. \name\ controls bias voltages of varactor diodes dynamically to mediate the propagation characteristics of adjacent media and boost transmission through the media boundary. It thus achieves the effect of ``medium virtualization'', i.e., the metasurface 
masks the fact that signals traverse a physical boundary, such that two (or more) media appear to be a single medium virtually.


The design for \name\ addresses several challenges.
First, we need to design a metasurface that can achieve media impedance matching. 
Second, our system needs to adapt to environmental dynamics, such as varying gaps between the metasurface and the media interface, different media, and changing multi-path channel conditions. 
Our control algorithm addresses all the adaptation needs simultaneously.
Lastly, the surface needs to be simple, lightweight, and conform to various shapes of the media interfaces. For example, for air-tissue networking, surfaces can be attached on fabric as part of a blanket or a hospital gown for medical applications. It is important to minimize the burden on users and also conform to the shape of human body.



We fabricated \name\ prototypes on thin, flexible PCB substrate (\secref{sec:impl}) together with a control circuit for voltage control. 
Extensive evaluation (\secref{sec:eval}) shows \name\ effectively enhances cross-media links.
\name\ provides over 10~dB reflection reduction for water and tissue, showing effective media impedance matching, thus increasing the received signal strength for air-to-water and air-to-tissue links.
For example, we provide a median of 55\% throughput increase and up to $4\times$ for air-to-water Wi-Fi links. 
For (emulated) in-water backsactter links, \name\ provides a median received power gain of 10~dB and up to 30~dB. Our system is also robust to changes to the surface-media gap or surface-endpoint distance.

In summary, this paper makes the following contributions. 
First, we highlight a new modality of signal manipulation for cross-media wireless networking via impedance matching of media. 
Second, we propose the first programmable metasurface design and an end-to-end system that can dynamically program reflection and transmission at media interfaces. It includes a novel metasurface hardware design and an efficient algorithm that dynamically achieve media impedance matching and beamforming control.
Third, we present a soft and lightweight metasurface implementation that facilitates flexible deployment. Experiments show that we can effectively mask the presence of a physical media boundary with \name.
%
This work does not raise any ethical issues.

%% file: motivation.tex
\section{Background and Motivation}
\label{sec:motivation}

We first explain RF signal propagation behavior across any media interface. 
Then we summarize existing solutions and their shortcomings to motivate our design. 

\begin{table}
    \caption{Different media show vastly different intrinsic impedance relative to air. \textmd{Any impedance mismatch causes strong reflection and weak transmission at media interfaces. \label{table:medium}}}    
        \small
    \begin{tabular}{m{2cm} || m{0.7cm}| m{0.7cm}| m{0.7cm}| m{0.7cm}| m{0.7cm}} 
        \hline
        Medium & Air & Water & Skin & Fat & Muscle\\
        \hline
        Relative permittivity &1 & 81 & 43.75 & 5.46 & 55.03 \\ 
        \hline
        Intrinsic impedance ($\Omega$) & 376.7 & 41.86 & 57.0 & 161.2 & 50.8\\
        \hline
    \end{tabular}

\end{table}


\subsection{Signal Propagation at Media Interface}
\label{sec:motiv:wave}

We characterize a medium with two parameters: \textit{relative permittivity} $\epsilon_r$ and \textit{relative permeability} $\mu_r$.
They capture how electromagnetic waves propagate in the medium, relative to vacuum.
Using these, we can derive the medium's intrinsic wave impedance $Z = \frac{\sqrt{\mu_r}}{\sqrt{\epsilon_r}} Z_0$, where $Z_0$ is the impedance of free space (vacuum).
The wave impedance describes the ratio of magnetic and electronic fields~\cite{microwave-engineering}.
Permeability equals 1 for all media we consider, so permittivity is the key parameter and decides the impedance of medium.

Assume signals travel from the air to another medium and they have relative permittivity $\epsilon_0=1$ and $\epsilon_r$ respectively.
Let $Z_0$ and $Z$ be the impedance of air and the medium of interest, solving the boundary conditions at the interface~\cite{microwave-engineering} gives
\begin{equation}
    \begin{aligned}
        \Gamma & = \frac{Z - Z_0}{Z+Z_0} = \frac{\sqrt{\epsilon_0} - \sqrt{\epsilon_r}}{\sqrt{\epsilon_0} + \sqrt{\epsilon_r}}\\
        T & = 1+\Gamma = \frac{2 Z}{Z+Z_0} = \frac{2 \sqrt{\epsilon_0}}{\sqrt{\epsilon_0} + \sqrt{\epsilon_r}}
    \end{aligned}
    \label{equ:reflex-at-interface}
\end{equation}
where $\Gamma$ and $T$ are the reflection and transmission coefficients of the interface respectively. The through-interface power is $|T|^2 \frac{Z_0}{Z}$, while the power reflected from the interface is $|\Gamma|^2$.
When the impedance of media are not equal or matched, there will be signal reflected and less power going through the interface.
Such power loss at the interface happens for both propagation directions.

\tabref{table:medium} lists the parameters of a few media according to ~\cite{yang_near-reflectionless_2022,DK-bio-tissue, EM-efficiency-bio}. 
We run HFSS~\cite{hfss} simulations with these parameters and the results are shown in \figref{fig:motiv-trans-reflex-comparison}.
The exact parameters may vary depending on the specific medium and measurement method, but small variations of such parameters do not change the results significantly.
We note that water and tissue (skin, fat, muscle) have much larger permittivity, which leads to strong reflection and weak transmission.
Tissue is a layered composition of various media, including skin, fat, muscle, thus there are multiple reflections until signals reach the in-vivo endpoint.

\heading{Attenuation and refraction.}
Media, like tissue and water, can have high conductivity, causing large attenuation. 
Conductivity also leads to a complex-valued permittivity.
For the media considered in the paper, the conductivity is within the range that does not affect impedance significantly.
In the following sections, we run most simulations with no tissue conductivity to isolate the performance of media impedance matching and show simulation results with conductivity in \figref{fig:design-hfss-gain-over-depth}. 
The experimental evaluation naturally includes the influence of conductivity and provides results considering all aspects.
Another phenomenon at the media interface is signal refraction, i.e., changing the propagation direction.
This can be viewed as contributing to multipath fading in the channel.

\heading{Backscatter links.}
State-of-the-art systems for in-vivo applications~\cite{yu_magnetoelectric_mobicom22,inNout-mobicom20, umedic-mobicom20, remix-sigcomm18,ivn-sigcomm18, inter-tech-backscatter} leverage backscatter communication to achieve battery-less networking. 
Backscatter links require both enough power harvested at the in-vivo endpoints and a sufficient SINR of received backscatter signal at the in-air endpoint.
The former suffers significant loss due to the weak transmission discussed in \secref{sec:motiv:wave}.
For the latter, received backscatter signals incur power loss (dB) twice at the media interface since the signals cross the interface twice.
The strong reflection at the interface also creates interference that lowers received SINR further.
Thus, the problem of media interface is more severe for backscatter links.


\subsection{Existing solutions and limitations}
\label{sec:motiv-exisitng}

\heading{Enhancing endpoints.}
IVN~\cite{ivn-sigcomm18} and In-N-Out~\cite{inNout-mobicom20} compensate for power loss with multiple transmitting antennas and beeamforming, but they rely on specialized and synchronized multi-antenna hardware. 
Such a hardware deployment can be costly and bulky, thus limiting practical usage.
Remix~\cite{remix-sigcomm18} and umedic~\cite{umedic-mobicom20} improves the in-vivo endpoint design. 
Such endpoint based solutions can not alter the RF propagation environment given where they reside, thus the best they can do is to mitigate the power loss caused by the media interface at the cost of hardware complexity.

\heading{Attaching antennas to the interface.} 
Some work propose specially designed, bio-matched antennas~\cite{theoretical_BMA_2020,BMA_low_2020, BMA_conformal_2017} that can achieve high efficiency when pressed against the skin. But this requires direct contact between endpoint and the medium, i.e., the user's skin. This brings discomfort and limits user mobility, thus, hinders long-term monitoring. Deploying an active endpoint close to the interface precludes using existing infrastructure such as Wi-Fi APs as the signal source, which likely requires battery power. By contrast, metasurfaces are passive and low-power, enhancing power from external signal sources.

\heading{Anti-reflection metasurfaces.} Recent studies propose to deploy metasurfaces~\cite{antireflection_cascaded,yang_near-reflectionless_2022, meta_matching_layer, Huygens_matching, milli_antireflection,air_ground_match_2020} at media interfaces to mitigate reflection.
Their designs, however, do not provide programmability, so the surface can not adapt to the changing environment. Nor are they able to cope with multipath fading channels.
Further, they rely on multiple layers of thick and rigid dielectric substrate to host the metasurface. Both limit their practical deployment. A rigid design also means they can not conform to the shape of media interface, e.g., human body. Lastly, they primarily focus on theoretical analysis, lacking an end-to-end system design and experimental validation.

\heading{Our approach.} 
In contrast to the above solutions, we present a flexible single-layer proragmmable media matching metasurface with beamforming capability. Matching the impedance of two adjacent media addresses the root cause of the reflections. We do not directly address signal attenuation or refraction, but viewed these via the perceived channel conditions at the receiver and use beamforming to mitigation these effects.
Note that our approach is complementary and orthogonal to existing efforts on improving \textit{endpoints}, thus not directly comparable. \name\ can be deployed alongside enhanced endpoints. 



%% file: design.tex
\section{\name{} Design}
\label{sec:design}
We aim to design programmable metasurfaces that mask the difference of media impedance.
We start with a circuit model analysis to understand how the surface should work. 
Based on the analysis, we set the design goals and emphasize the importance of programmability.
Next, we study the design of programmable metasurface element pattern with a large tuning range.
On that basis we design an algorithm to control the surface efficiently. 


\begin{figure*}[t]
    \centering
    \begin{subfigure}[b]{0.3\textwidth}
        \includegraphics[width=0.9\columnwidth]{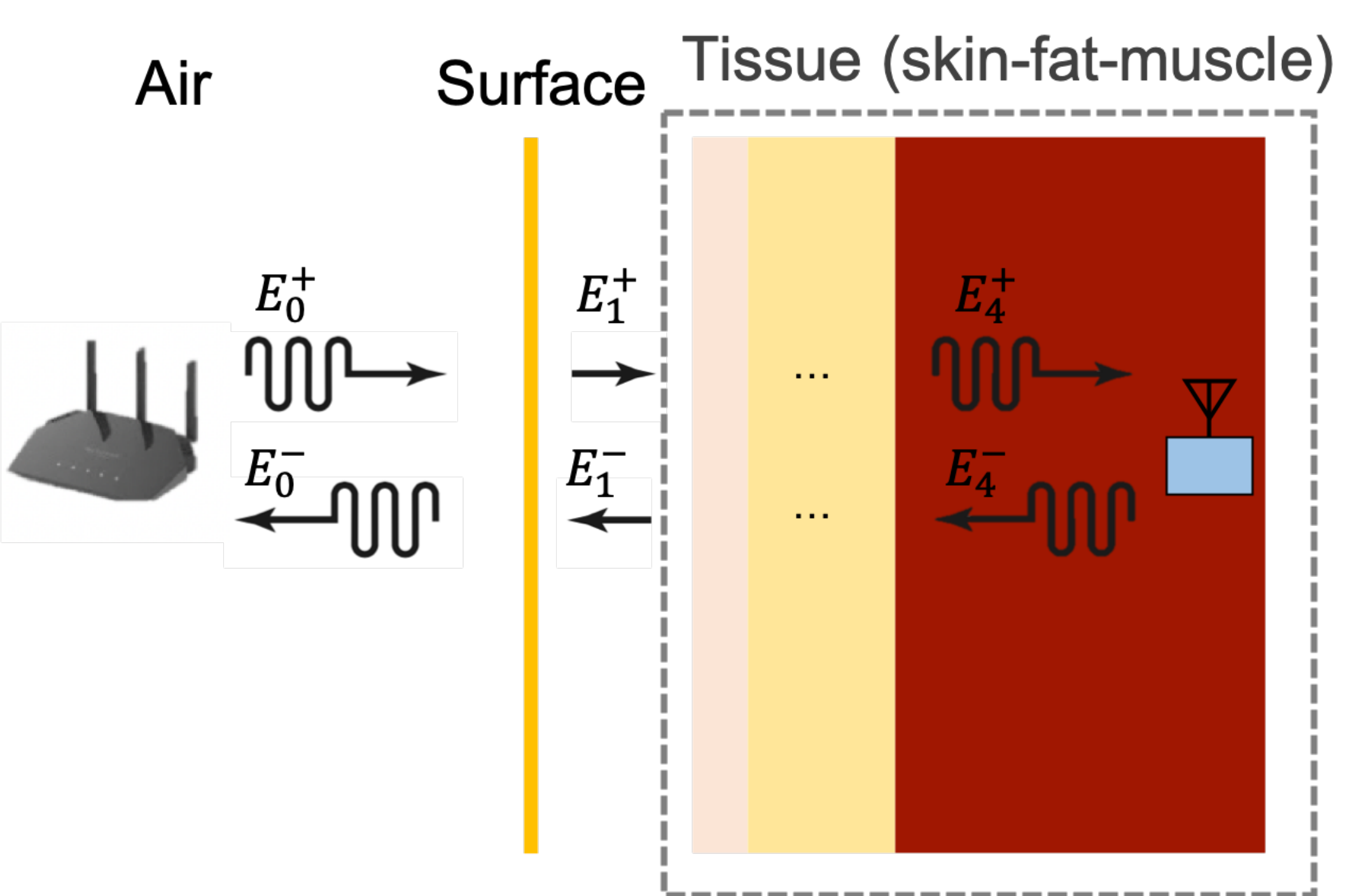}
        \caption{}
        \label{fig:design-tissue-model}
   \end{subfigure}\hspace{1em}%
   \begin{subfigure}[b]{0.48\textwidth}
        \includegraphics[width=0.9\columnwidth]{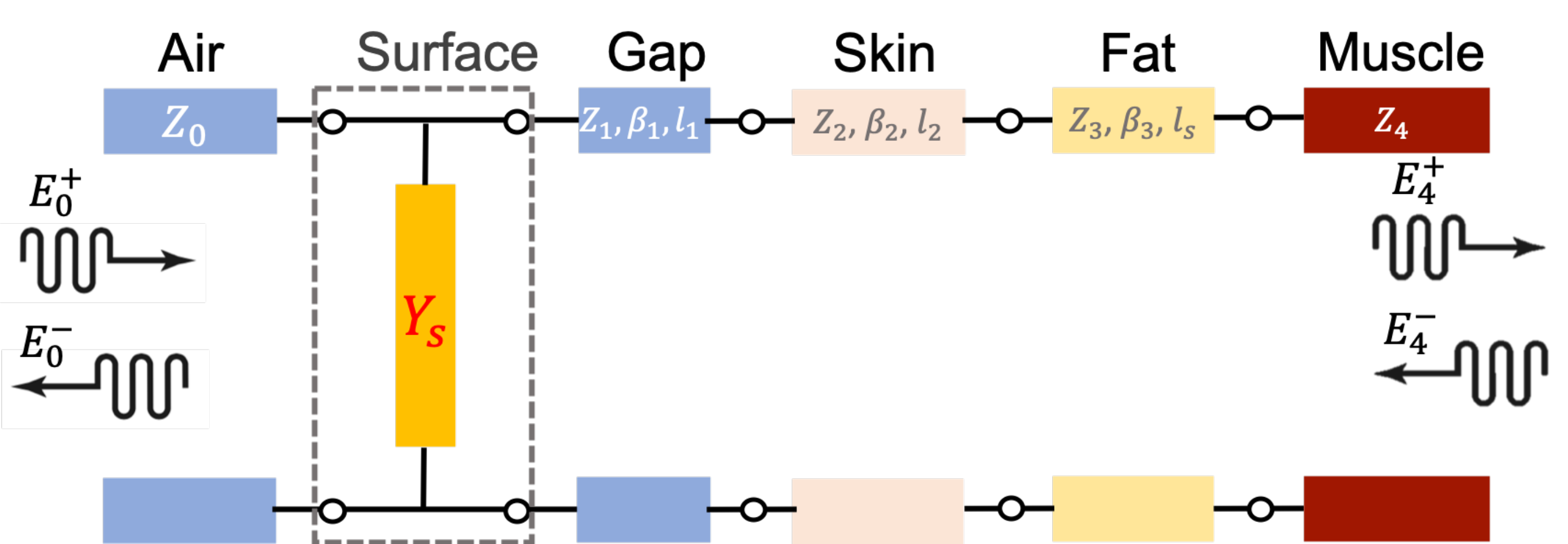}
        \caption{}
        \label{fig:design-tissue-circuit}
    \end{subfigure}
   \begin{subfigure}[b]{0.18\textwidth}
      \includegraphics[width=0.95\columnwidth]{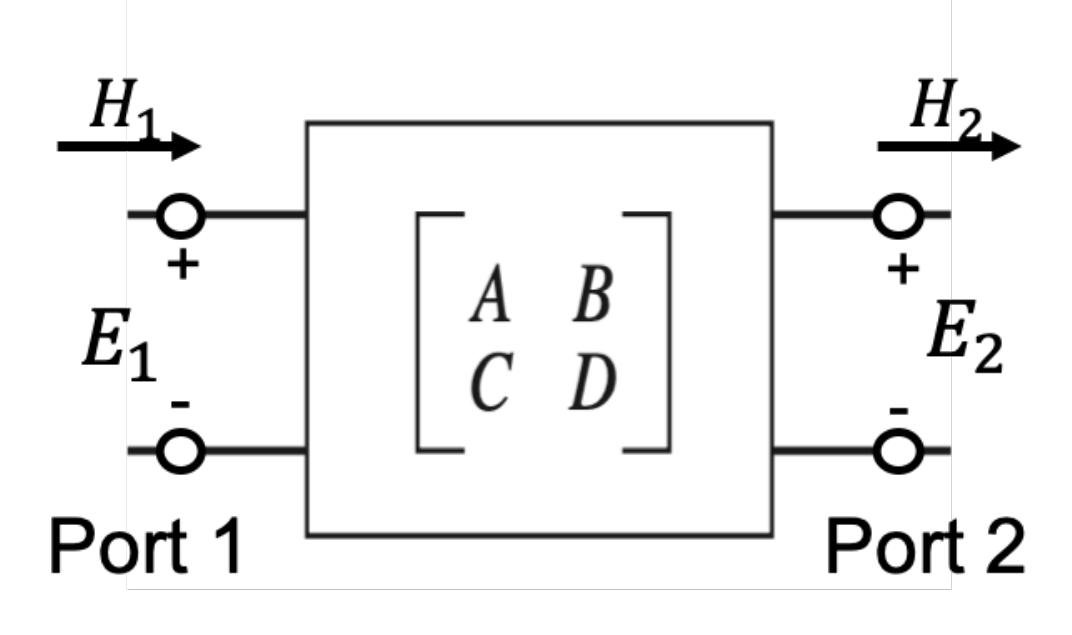}
      \caption{}
      \label{fig:design-ABCD-def}
   \end{subfigure}\hspace{1em}%
   \caption{Analysis of surface operations. \textmd{\textbf{(a)} Analytical model of surface and tissue layers. \textbf{(b)} Circuit model of surface and tissue layers. \textbf{(c)} The definition of ABCD wave matrix. We model the propagation of waves with a cascaded two-port microwave network to understand the needed \textit{surface admittance ($Y_s$)} to minimize reflection $E_0^{-}$ and boost through-interface transmission $E_4^{+}$.}}
   \label{fig:motiv-sim-tunability}
 \end{figure*}

\subsection{Operational Principle}
\label{sec:design-principle}

We analyze how a metasurface alters wave propagation between air and tissue to guide our system design. Tissue is the most complex case due to its layered structure of multiple substances. Other media can be modeled similarly.



\heading{Surface at the media interface.}
\figref{fig:design-tissue-model} shows the analytical model for RF signals propagating through air-tissue interface, where the tissue is a layered composition of skin, fat and muscle.
The wireless endpoint on the left transmits RF signals towards the surface and the interface.
After going through the surface and any potential gaps, signals propagate through tissue layers.
We do not assume a fixed surface-media gap or the specific material in between.
The gap can be filled with air or fabric or other dielectric materials depending on the application scenarios.
This way, we decouple the surface and media, relaxing the requirement of surface deployment.
The signals arrive at the in-vivo endpoints surrounded by muscle or other organ.
Based on the plane wave solution~\cite{microwave-engineering}, the total electronic field in each medium contains forward propagating $E^+$ and backward propagating $E^-$ waves, i.e., $E = E^+ + E^-$.
We want to minimize the reflection $E_0^-$ from the interface and maximize the through-interface transmission $E_4^+$ towards the in-vivo endpoint.

\heading{Characterization as circuit components.}
Although we described how waves propagate in \secref{sec:motiv:wave},
the problem gets complex with multiple layers of media and the metasurface.
Waves resonate, i.e., bounce back and forth, between different tissue layers and the surface. 
They also experience various phase delays related to media thicknesses.
To characterize the problem, we consider the microwave circuit model shown in \figref{fig:design-tissue-circuit}.
Each segment of medium is considered as a transmission line, characterized by its intrinsic impedance $Z$, phase constant $\beta$ and length $l$. 
The metasurface operates by the surface current induced by the incident signals, thus it can be characterized as a shunt circuit component with \textit{\textbf{Surface Admittance $Y_s = G + jB$}}, as per existing work~\cite{antireflection_cascaded, yang_near-reflectionless_2022, cascaded_complete_control}.
It is a complex value, with a real part (the \textit{conductance}, $G$) and an imaginary part (the \textit{susceptance}, $B$).
A lossless surface should have a small or no conductance, so we adjust the imaginary part for media impedance matching.
The surface hardware design determines the exact value of $Y$, characterizing how signals interact with the surface quantitatively.

\heading{Cascading all components.}
To analyze the interactions among the surface and the media, we cascade the afore-mentioned circuit components, each as a two-port microwave circuit network.
Thus, the circuit model is a cascade of two-port networks as in the \figref{fig:design-tissue-circuit}, where ports are marked with circles.
Then, we use a 2x2 transmission matrix, i.e., ABCD matrix, to represent each two port network.
For the whole cascaded circuit model, the matrix is written as 
\begin{equation}
    {\left[\begin{array}{c}
    E_0 \\
    H_0
    \end{array}\right] } =\left[\begin{array}{ll}
    A & B \\
    C & D
    \end{array}\right]\left[\begin{array}{c}
    E_4 \\
    H_4
    \end{array}\right]
    \label{equ:ABCD-EMfields}
\end{equation}
We find the ABCD matrix by multiplying known matrices of the surface and the media, which are a shunt circuit component and transmission lines~\cite{microwave-engineering}.
\begin{equation}
    \left[\begin{array}{ll}
    A & B \\
    C & D
    \end{array}\right]=
    \left[\begin{array}{cc}
    1 & 0 \\
    Y_s & 1
    \end{array}\right]
    \prod_{k=1}^4\left[\begin{array}{cc}
    \cos \beta l_k & j Z_k \sin \beta l_k \\
    j \sin \beta l_k / Z_k & \cos \beta l_k
    \end{array}\right]
    \label{equ:ABCD-multipy}
\end{equation}
Then, we can solve the transmission and reflection coefficients based on ABCD matrix values (Appendix).
\begin{equation}
    \begin{aligned}
        T = \frac{E_4^{+}}{E_0^{+}} = \frac{2}{A+B/Z_4 +CZ_0+DZ_0/Z_4}\\
        \Gamma = \frac{E_0^{-}}{E_0^{+}} = \frac{A+B/Z_4 -CZ_0-DZ_0/Z_4}{A+B/Z_4 +CZ_0+DZ_0/Z_4}
    \end{aligned}
    \label{equ:S-from-ABCD}
\end{equation}
Compared to \autoref{equ:reflex-at-interface}, after deploying the surface, the propagation behavior at the interface is no longer solely decided by the media impedance.

\heading{Summary.}
To summarize, the metasurface alters signal propagation with its \textit{surface admittance}, showing by the cascaded ABCD matrix \autoref{equ:ABCD-multipy}. Therefore, it can affect how waves propagates through the interface, i.e., change transmission and reflection coefficients in \autoref{equ:S-from-ABCD}. 
This offers us the opportunity to match the impedance of different media with a metasurface.

\heading{Other media interfaces.}
The air-water and air-ground interfaces can be modeled in the same way.
Since they involve matching only two media, these are simpler cases than air-tissue interfaces and the above analysis can be extended to these cases easily. 

\subsection{Design Goals}
Given the analysis above, we want to set the design goals by finding the appropriate surface admittances.

\heading{Single-layer surface.}
We consider one layer surface instead of a cascade of multiple surfaces, thus only one admittance value to decide.
This offers simplicity, low hardware cost, and ultra-thin thickness.
The optimal solution for media impedance matching can be derived analytically with 3 layers non-programmable metasurfaces~\cite{antireflection_cascaded}.
However, according to our calculation, one layer surface is already able to achieve near-optimal performance.

\begin{figure}[t]
    \centering
  \begin{subfigure}{0.33\columnwidth}
    \centering
    \includegraphics[scale=0.2]{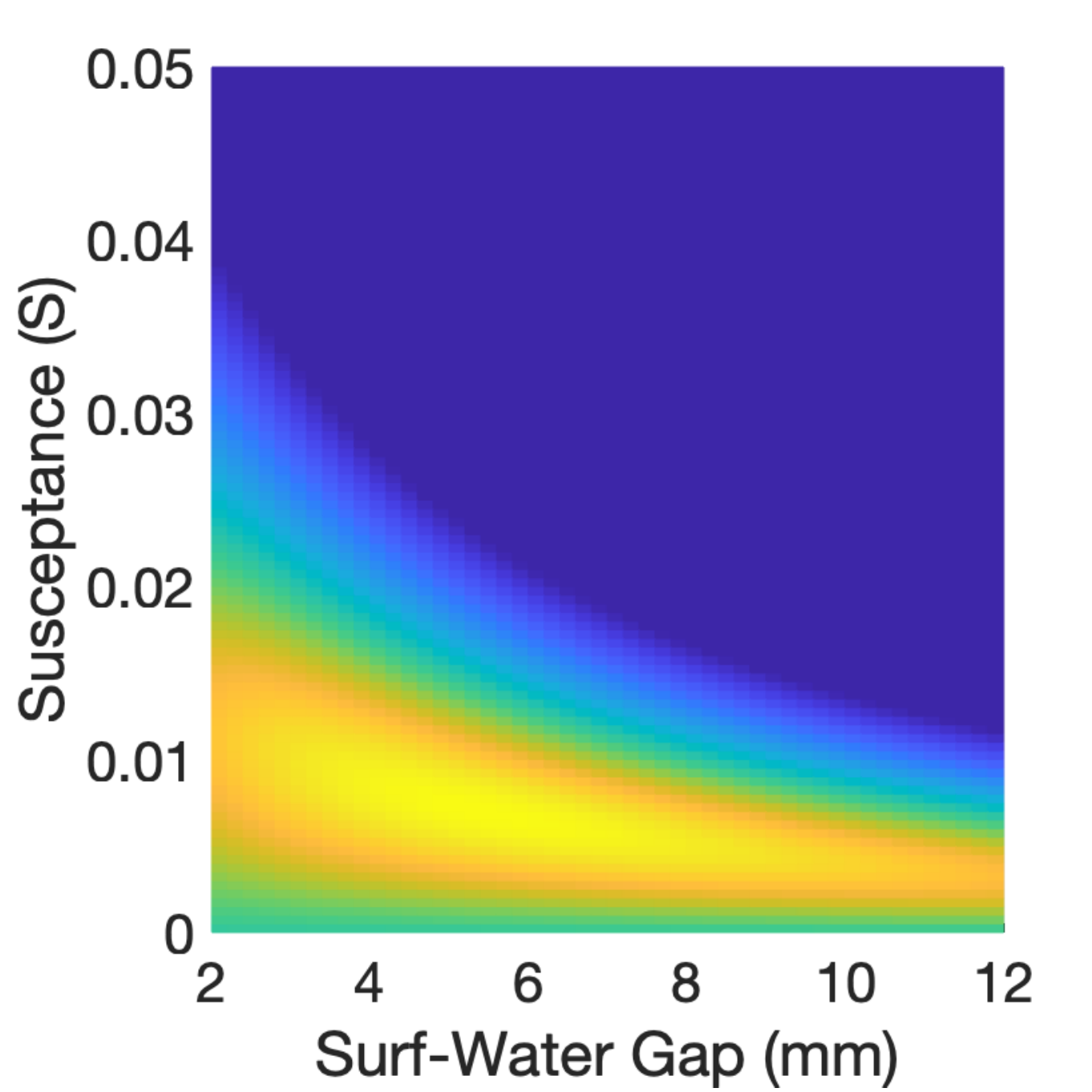}
    \caption{}
    \label{fig:design-theory-water-gap-heatmap}
  \end{subfigure}%
  \begin{subfigure}{0.33\columnwidth}
    \centering
    \includegraphics[scale=0.2]{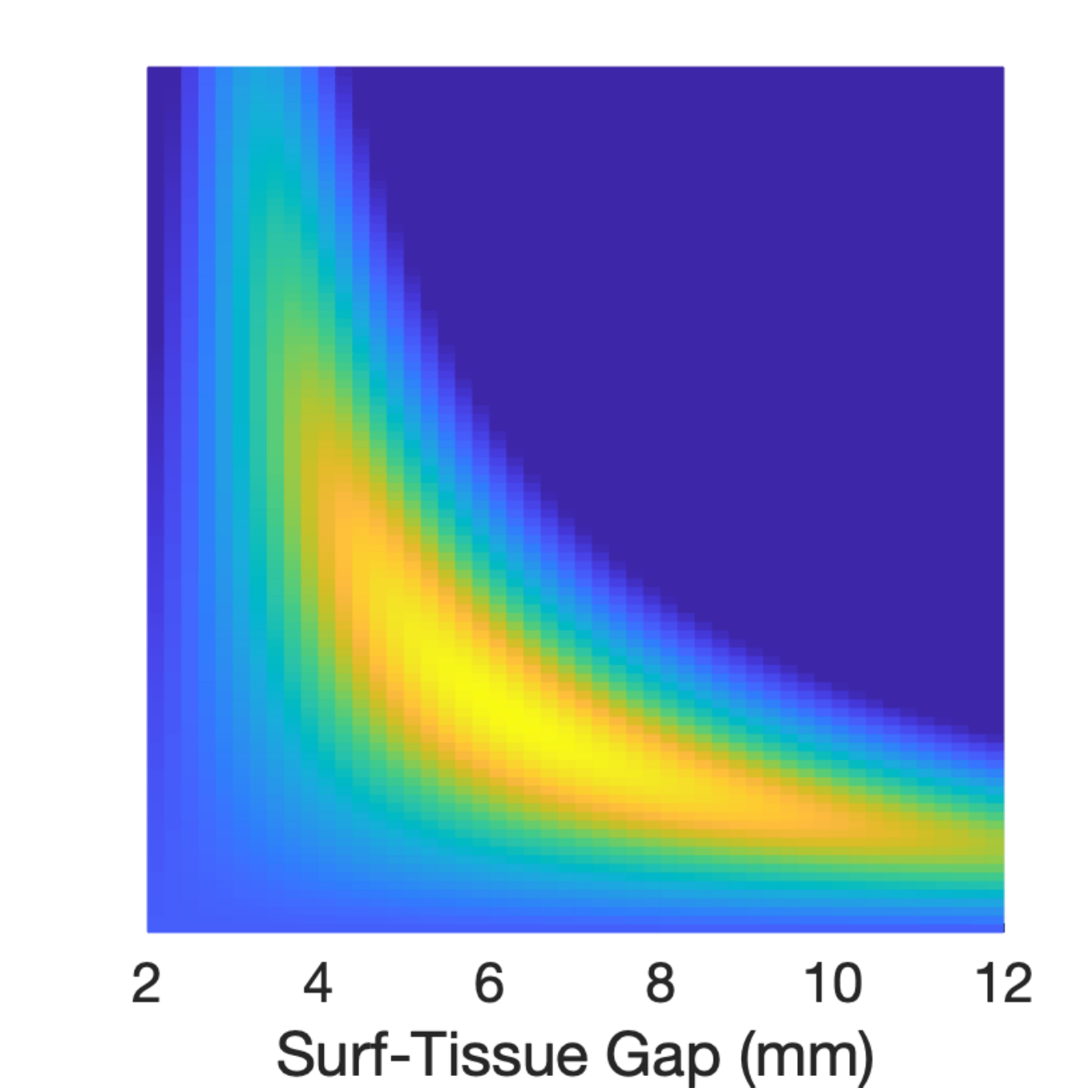}
    \caption{}
    \label{fig:design-theory-tissue-gap-heatmap}
  \end{subfigure}
  \begin{subfigure}{0.33\columnwidth}
    \centering
    \includegraphics[scale=0.2]{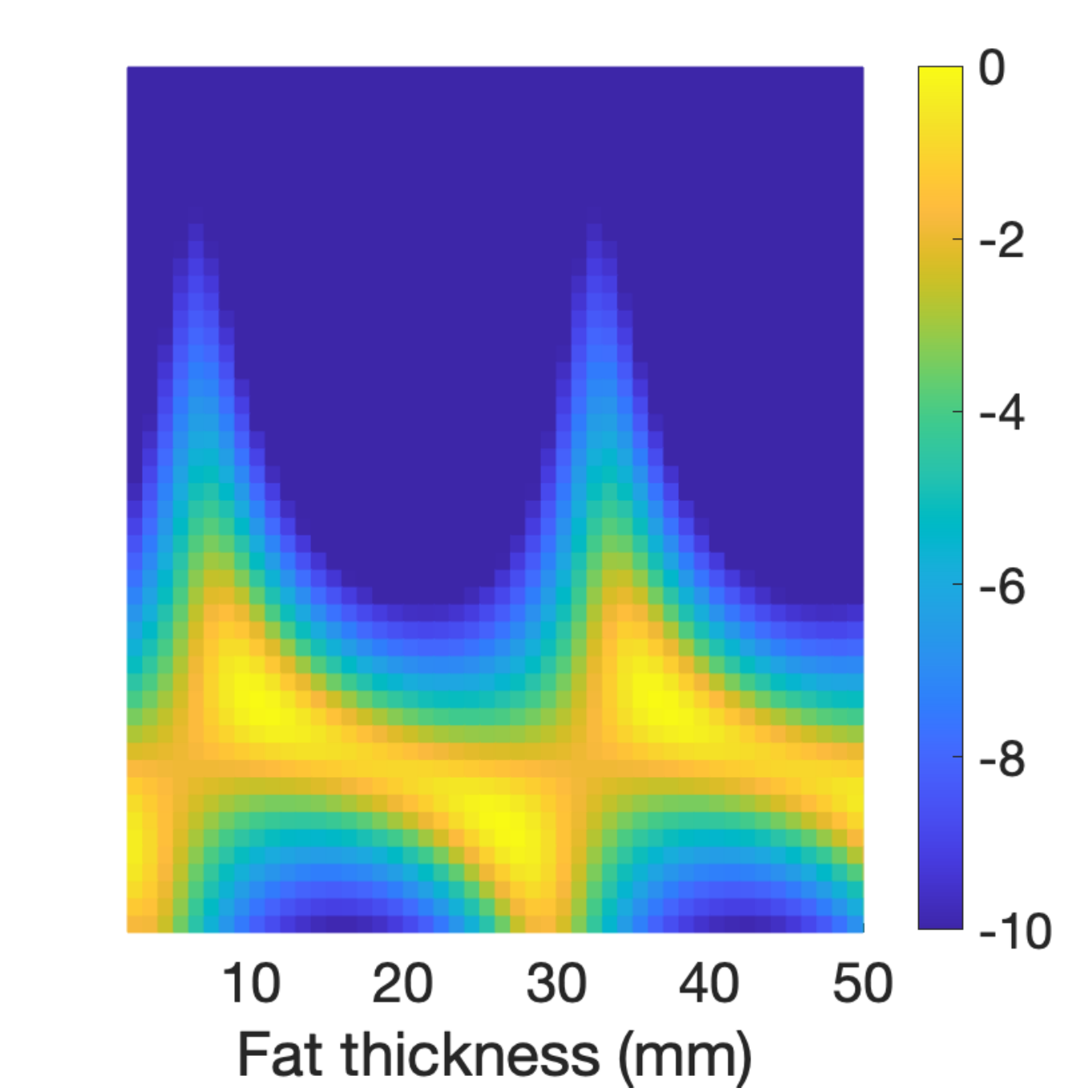}
    \caption{}
    \label{fig:design-theory-tissue-fat-heatmap}
  \end{subfigure}
  \caption{Heatmaps of through-interface power versus surface admittance. \textmd{The desired surface admittance changes with \textbf{(a) (b)} different surface-media gap distances or \textbf{(c)} thickness of fat. This highlights the necessity of surface programmability.} }
  \label{fig:design-theory-heatmap}
\end{figure}

\heading{Search for admittance.}
We numerically search for the appropriate surface admittance to match the media, i.e., minimizing reflection and maximizing transmission.
This is a one-time offline search process for the hardware design. 
For any given surface admittance $Y_s$, we can calculate the ABCD matrix based on \autoref{equ:ABCD-multipy}. 
Then, we calculate the transmission and reflection coefficients based on \autoref{equ:S-from-ABCD}.
\figref{fig:design-theory-heatmap} shows the power transmitted through the media interface when using different surface admittance and surface-media gap distances or fat thicknesses. The higher the through-interface power is, the lower the reflection is, and thus, the better the media matching performance. 
We can achieve near-optimal transmission (only 0.5~dB loss when transmitting through the interface) by using the appropriate surface admittance.

\heading{Operational frequencies.}
Several proposals~\cite{inter-tech-backscatter, FSbackscatter, backfi} demonstrate the possibility of exploiting 2.4~GHz Wi-Fi/bluetooth radios as ambient signal sources to provide backscatter connectivity to implanted radios. 
This could significantly lower the hardware requirement for in-vivo networking and make it more accessible for daily use. 
Also the antenna size is inversely proportional to the frequency given the same performance requirement, and a small antenna size is important for in-vivo applications.
Thus we focus on 2.4~GHz in this paper. The design principle shown can be applied to other frequencies like 900~MHz straightforwardly.


\heading{The need for programmability.}
However, there is no one-size-fits-all surface admittance. 
Several factors can affect the appropriate surface admittance for impedance matching:
(\romannumeral 1) the gap between surface and medium affects the needed admittance significantly. 
As shown in \figref{fig:design-theory-water-gap-heatmap} and \figref{fig:design-theory-tissue-gap-heatmap}, 
when the gap gets smaller, we need a larger admittance to maintain a high through-interface transmission power.
For a practical deployment, we can hardly control the gap size precisely. 
Thus, we aim to adapt to the gap with a programmable surface.
(\romannumeral 2) Different kinds of media or media with different compositions need different admittances. 
\figref{fig:design-theory-water-gap-heatmap} and \figref{fig:design-theory-tissue-gap-heatmap} show that water need a much lower admittance than tissue when the gap is small.
Another example is that the thicknesses of fat layers vary significantly from one person to another. 
Because of the small permittivity of fat~(\tabref{table:medium}), its thickness can influence the admittance we need significantly as shown in \figref{fig:design-theory-tissue-fat-heatmap}.
A surface with a large admittance tuning range can be a more general hardware design for various deployment scenarios.
(\romannumeral 3) We build the circuit model with an approximate tissue model and simulate designs with lab measured medium parameters, both simplified from real scenarios. 
We can use programmability to adapt to any unaccounted inaccuracy.

Thus, our goal is to design a one-layer programmable metasurface covering a sufficient surface admittance range, e.g., from $0j$ to over $0.1j$. 
This is challenging since drastically different surface patterns are often needed to account less than $0.01j$ change~\cite{yang_near-reflectionless_2022}.


\subsection{Hardware Design}
In this section, we present the design of metasurface hardware, including how we achieve programmability and low loss.

\begin{figure}[t]
    \centering
  \begin{subfigure}{0.45\columnwidth}
    \centering
    \includegraphics[scale=0.25]{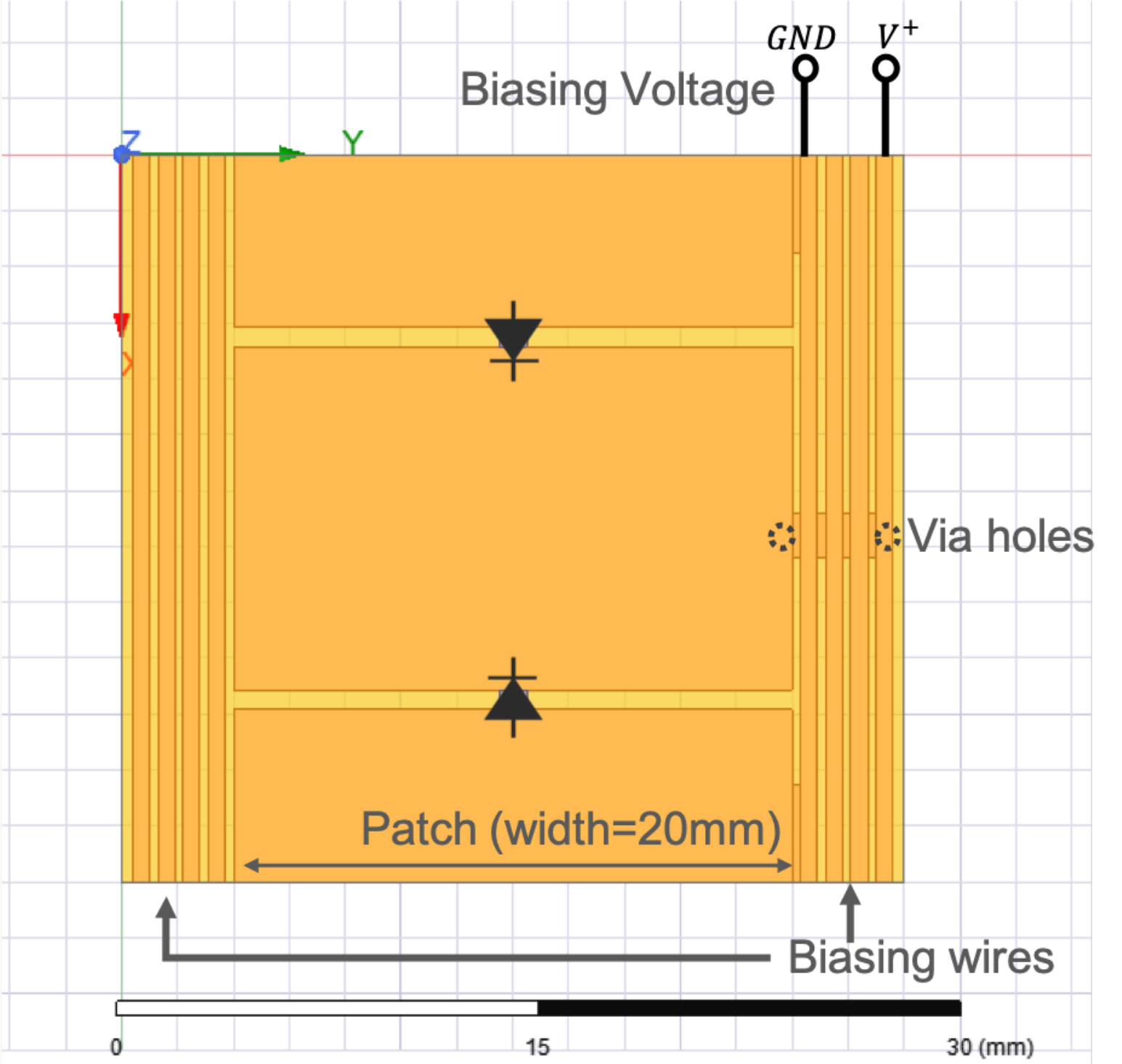}
    \caption{}
    \label{fig:design-element-pattern}
  \end{subfigure}%
  \begin{subfigure}{0.45\columnwidth}
    \centering
    \includegraphics[scale=0.4]{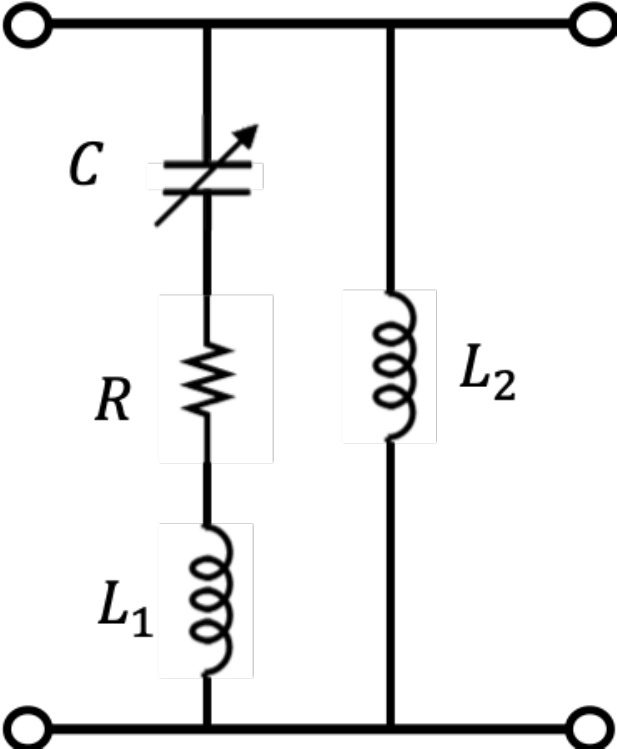}
    \caption{}
    \label{fig:design-element-circuit}
  \end{subfigure}
  \caption{Metasurface Pattern Design. \textmd{\textbf{(a)} Surface element pattern. \textbf{(b)} Equivalent LC circuit of the pattern .} \textmd{We control the voltage applied to varactor diodes with individual biasing wire for each element.} }
  \label{fig:design-element}
\end{figure}


\heading{Surface Pattern.}
We use a simple yet effective surface pattern, rectangular patches connected by varactor diodes, inspired by existing designs~\cite{wu_polar_rotator_2019, wu_space-time-surface_2020, llama} for other purposes.
The pattern is made of an array of \textit{elements}, each having two varactor diodes and associated patches as shown in
\figref{fig:design-element-pattern}.
The metallic patches act like antennas, creating necessary electronic coupling and inducing surface currents. The varator diodes work as tunable capacitors to alter the electronic response and produce the desired electronic surface admittance.
We add biasing wires next to patches to apply bias voltages to varactor diodes.
All surface elements share a common ground wire as voltage reference,
while each element has a separate biasing wire, connected through via holes and wires at back, for voltage control for element-wise control.
We use also such control to provide extra beamforming gain as shown in \figref{fig:eval-algo-cdf-control}.

We do not consider more sophisticated designs like Huygens metasurfaces~\cite{Huygens_matching, Huygens_surface_PRL} because the elements would need to be arranged parallel to the wave propagation direction, which means the thickness of surface depends on the wavelength, and can get very thick for 2.4~GHz or lower frequency.

\begin{figure}[t]
    \centering
    \includegraphics[width=0.7\columnwidth]{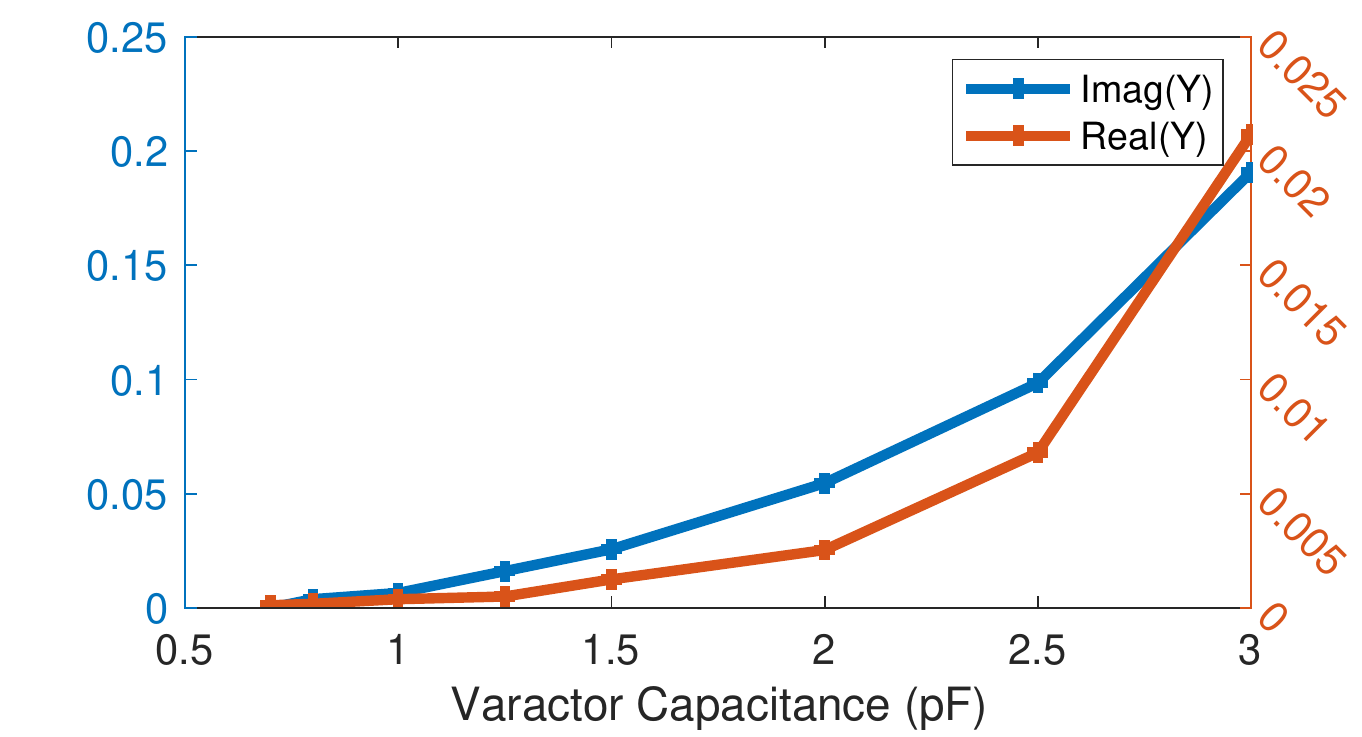}
    \caption{Surface admittance versus capacitance. \textmd{The surface shows an increasing admittance when the capacitance of varactor increases; the real part (lossy) of admittance is kept low.}}
    \label{fig:design-Y-over-cap}
\end{figure}

\heading{Varactors for programmability.}
Varactor diodes show different capacitance according to the biasing voltage applied.
The equivalent circuit for the surface is shown as \figref{fig:design-element-circuit},
where $C$ is the tunable capacitance from varactors and surface pattern, $R$ is the resistance of varactors,
$L_1$ is the inductance of patches, and $L_2$ is the inductance of biasing wires.
Based on this, we know surface admittance
\begin{equation}
    \begin{aligned}
        Y_s &= \frac{1}{\frac{1}{j\omega C} + R + j\omega L_1} + \frac{1}{j\omega L_2}\\
        &= \frac{\omega^2 C^2 R + j \omega C (1-\omega^2 C R)}{(1-\omega^2 C L_1)^2 + (\omega C R)^2} - \frac{j}{\omega L_2}\\
        &\approx \frac{\omega^2 C^2 R}{(1-\omega^2 C L_1)^2} + (\frac{\omega C}{1-\omega^2C L_1} - \frac{1}{\omega L_2})j
    \end{aligned}
    \label{equ:pattern-circuit}
\end{equation}
For our design, $(\omega C R)^2 \ll (1-\omega^2 C L_1) $, thus we make the approximation in the last step.
We can see from the equation that 
the tunable capacitance $C$ is the dominant factor for the susceptance (the imaginary part of admittance), which enables programmability.
The conductance (the real part of admittance) is proportional to resistance $R$, which causes additional power loss on the surface.
To minimize the lose and maximize tunability, we need to use varactors with small resistance but large capacitance tuning range.
The exact capacitance of varactors should also match the realizable value of patch inductance $L_1$ and biasing wire inductance $L_2$, so that the final admittance can be tunable in the wanted range, i.e. $0j$ to over $0.05j$.

\begin{table}
  \small
  \begin{tabular}{m{2.3cm} || m{0.5cm}| m{0.5cm}| m{0.5cm}| m{0.5cm}| m{0.5cm} | m{0.5cm}} 
      \hline
      Voltage (V) & 30 & 20 & 15 & 10 & 5 & 0\\
      \hline
      Capacitance (pF) &0.71 & 0.81 & 0.90 & 1.0 & 1.32 & 3.72\\ 
      \hline
      Resistance ($\Omega$) & 0.26 & 0.3 & 0.36 & 0.38 & 0.45 & 0.63\\
      \hline
  \end{tabular}
  \caption{Varactor capacitance and resistance verses reverse bias voltage.}
  \label{table:varactor}
\end{table}


\heading{Fine tuning the pattern.}
For our design, we choose SMV1405~\cite{SMV1405} varactor diodes. 
We run SPICE simulation to get the capacitance and resistance verses bias voltage (\tabref{table:varactor}). 
Provided the varactor and its capacitance range, we tune the pattern to achieve the desired admittance.
We guide the design with \autoref{equ:pattern-circuit}.
If we want to increase the ratio of susceptance and varactor capacitance, we can length of surface patch for higher $L_1$;
If we want to decrease the susceptance by a constant, we can increasing width of biasing wire for lower $L_2$.


\heading{Tunable admittance range.} 
Through iterative HFSS~\cite{hfss} simulations, we arrive at the final design shown in \figref{fig:design-element-pattern}.
We extract the surface admittance when varactor showing different capacitance in \figref{fig:design-Y-over-cap}.
The susceptance (imaginary part of admittance) changes from $0j$ to over $0.1j$, 
while the conductance (real part) keeps one order of magnitude lower than susceptance.
Thus, we achieve a large admittance tuning range but still keeps the loss low.
This way, we have a programmable metasurface for all situations instead of different non-progammable surfaces to achieve different admittance.


\heading{Media coupling.}
We note that the surface admittance may not always be a fixed function of capacitance/voltage.
Unexpected coupling between the surface and media can happen, especially when they are placed very close, and change the surface admittance. 
A thick PCB substrate can be used to reduce potential coupling, but will make the surface bulky and heavy.~\cite{yang_near-reflectionless_2022}.
Thanks to the programmability, our surface simply can adjust applied voltage to compensate any deviation caused by coupling.
We only need to make sure the tunable range is large enough to account for media coupling.
According to our simulation, admittance can get lower due to coupling with water or tissue.
But our surface can still perform media impedance matching well as shown next.

\begin{figure}[t]
  \centering
\begin{subfigure}{0.45\columnwidth}
  \centering
  \includegraphics[scale=0.3]{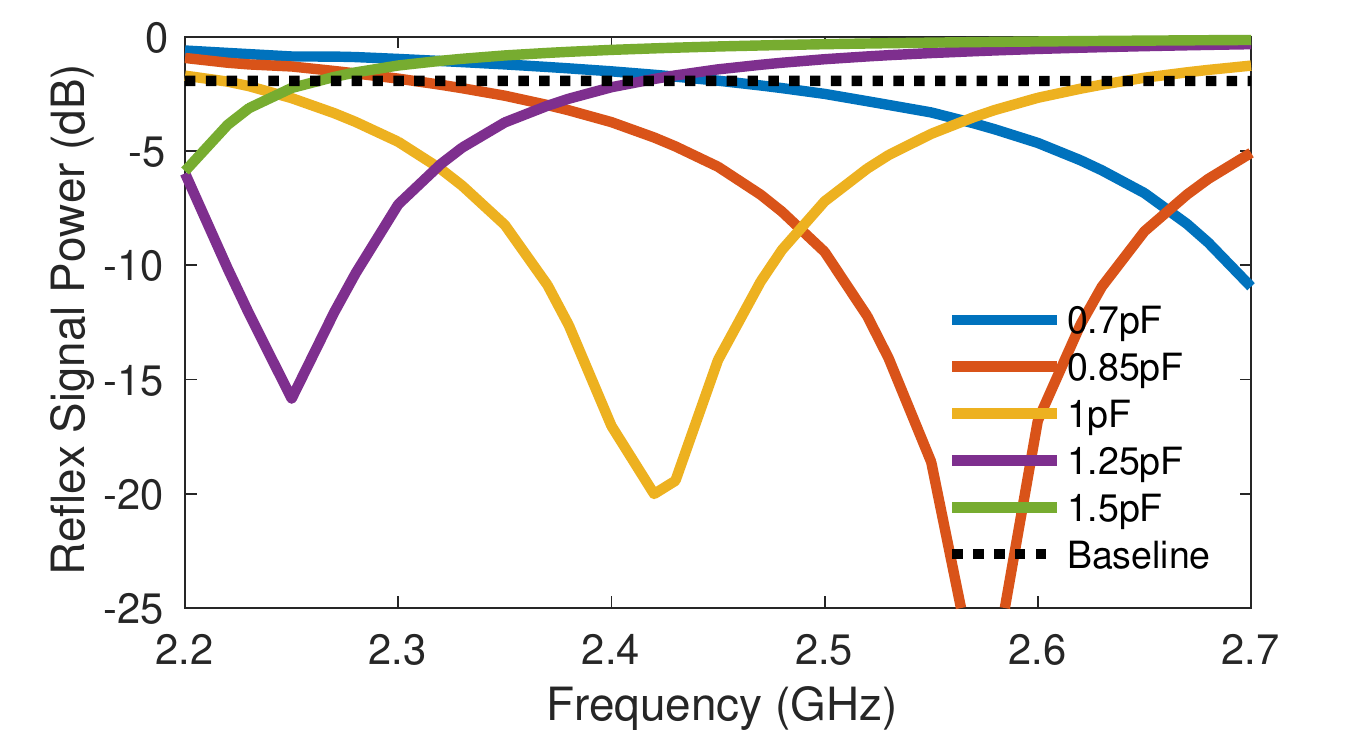}
  \caption{}
  \label{fig:design-hfss-reflex-water}
\end{subfigure}
\begin{subfigure}{0.45\columnwidth}
  \centering
  \includegraphics[scale=0.3]{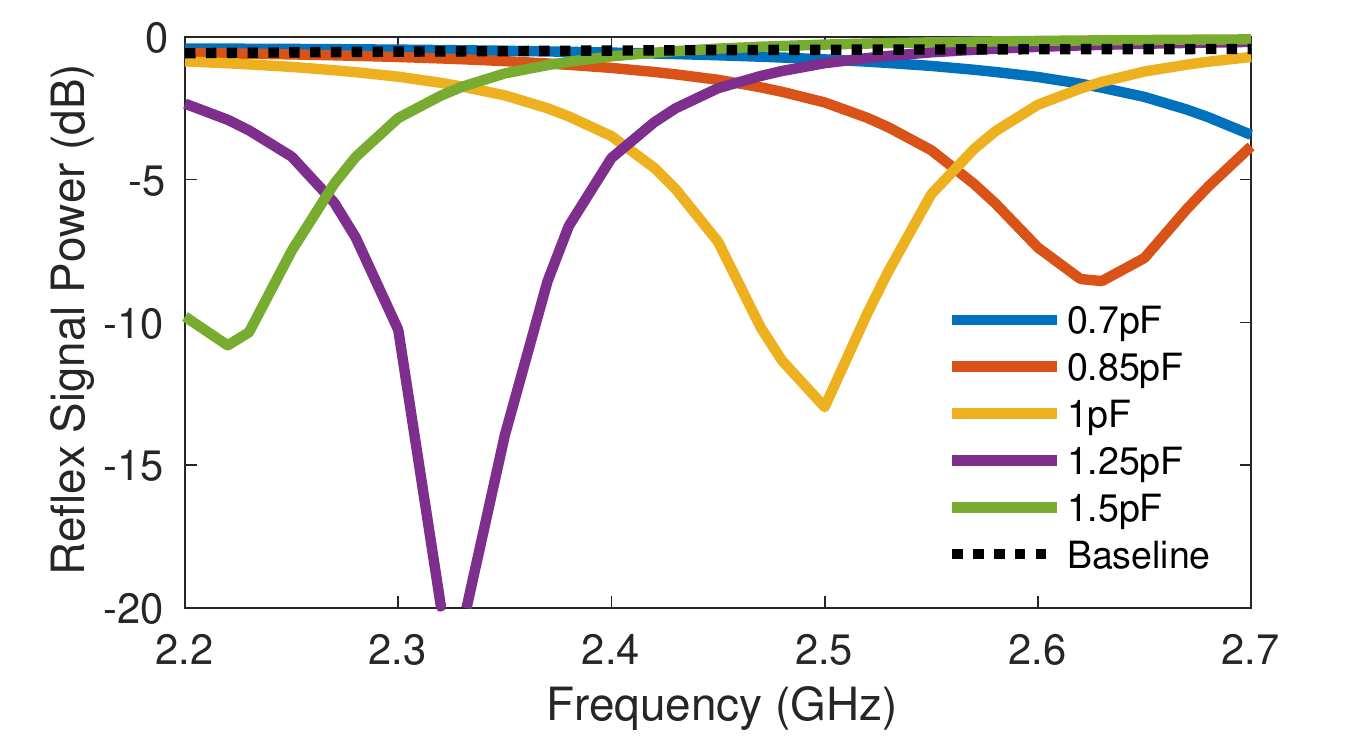}
  \caption{}
  \label{fig:design-hfss-reflex-tissue}
\end{subfigure}
\caption{Reflection reduction over frequency. \textmd{With appropriate capacitances, the surface achieves over 10~dB reflection reduction for water~\textbf{(a)} and tissue~\textbf{(b)} around 2.4~GHz.}}
\label{fig:design-hfss-reflex-over-freq}
\end{figure}

\begin{figure}[t]
  \begin{minipage}{.48\columnwidth}
    \centering
    \includegraphics[width=\columnwidth]{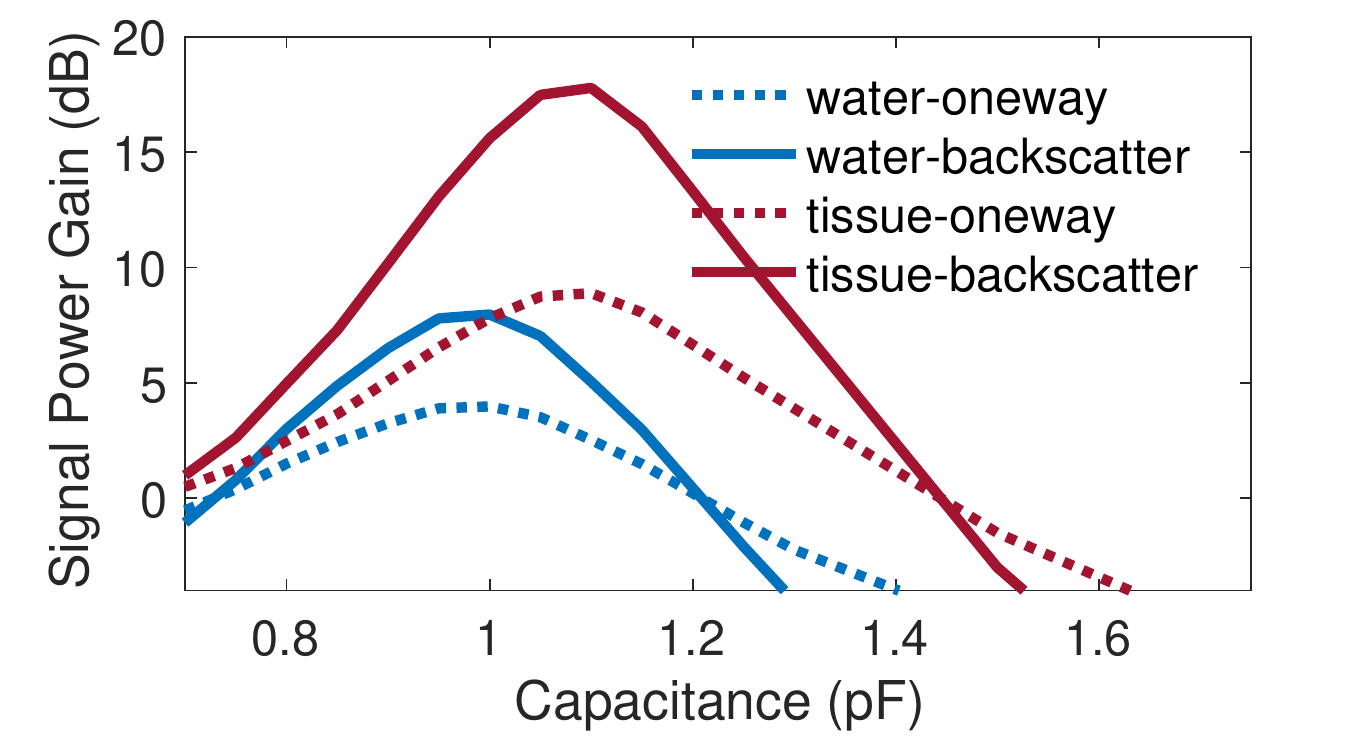}
    \caption{Oneway and Backscatter link gain.}
    \label{fig:design-hfss-backscatter-gain}
  \end{minipage}\hfill
  \begin{minipage}{.48\columnwidth}
    \centering
    \includegraphics[width=\columnwidth]{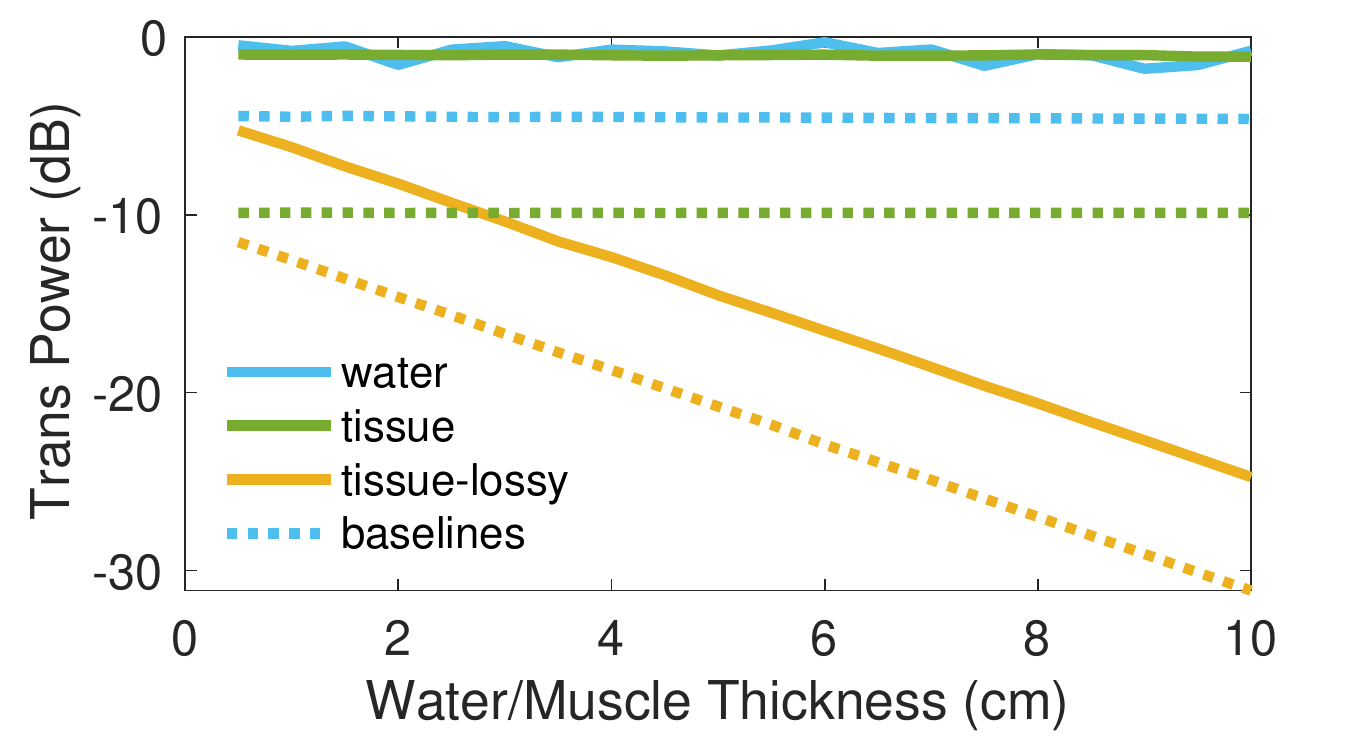}
    \caption{Transmission gain vs. endpoint depth.}
    \label{fig:design-hfss-gain-over-depth}
  \end{minipage}
\end{figure}

\heading{Media impedance matching simulations.}
To verify our design, we run HFSS~\cite{hfss} simulations for surfaces placed before air-tissue and air-water interfaces.
We first examine the change of reflection caused by the surface as shown in \figref{fig:design-hfss-reflex-over-freq}.
If appropriate capacitances are chosen, around 1~pF, our surface can achieve over $-10$~dB reflection reduction for both water and tissue, with a sufficient bandwidth around 2.4~GHz.
The reflection reduction trough deviates to neighboring frequencies when the capacitance is slightly different.
The transmission power through the interfaces with our surface show the reversed trend of reflection over frequency.
So we focus on the transmission power gain versus different capacitances, as shown in \figref{fig:design-hfss-backscatter-gain}.
Our surface provides 4~dB and 9~dB transmission gains for water and tissue respectively.
Almost all power are transmitted through in both cases, but the air-tissue interface has a worse baseline power, thus, we observe more gain.
It is worth noticing that our surface provides equal gains for both propagation directions.
Assuming the backscattered signal strength is proportional to the received signal strength, the gain for backscatter signal is the multiplication of gain of two directions (i.e., addition of the gain values in dB).
We achieve up to 8~dB and 18~dB gain for backscatter devices in water and tissue respectively by matching the media impedance.
Next, we simulate the influence of endpoint in media depth and media conductivity in \figref{fig:design-hfss-gain-over-depth}.
For both water and tissue, the transmission gain stays constant regardless of the depth.
If we consider the conductivity of tissue, the received signal power decreases when depth grows,
but our surface still provides a high gain regardless of depth.

\begin{figure}[t]
  \centering
\begin{subfigure}{0.33\columnwidth}
  \centering
  \includegraphics[scale=0.2]{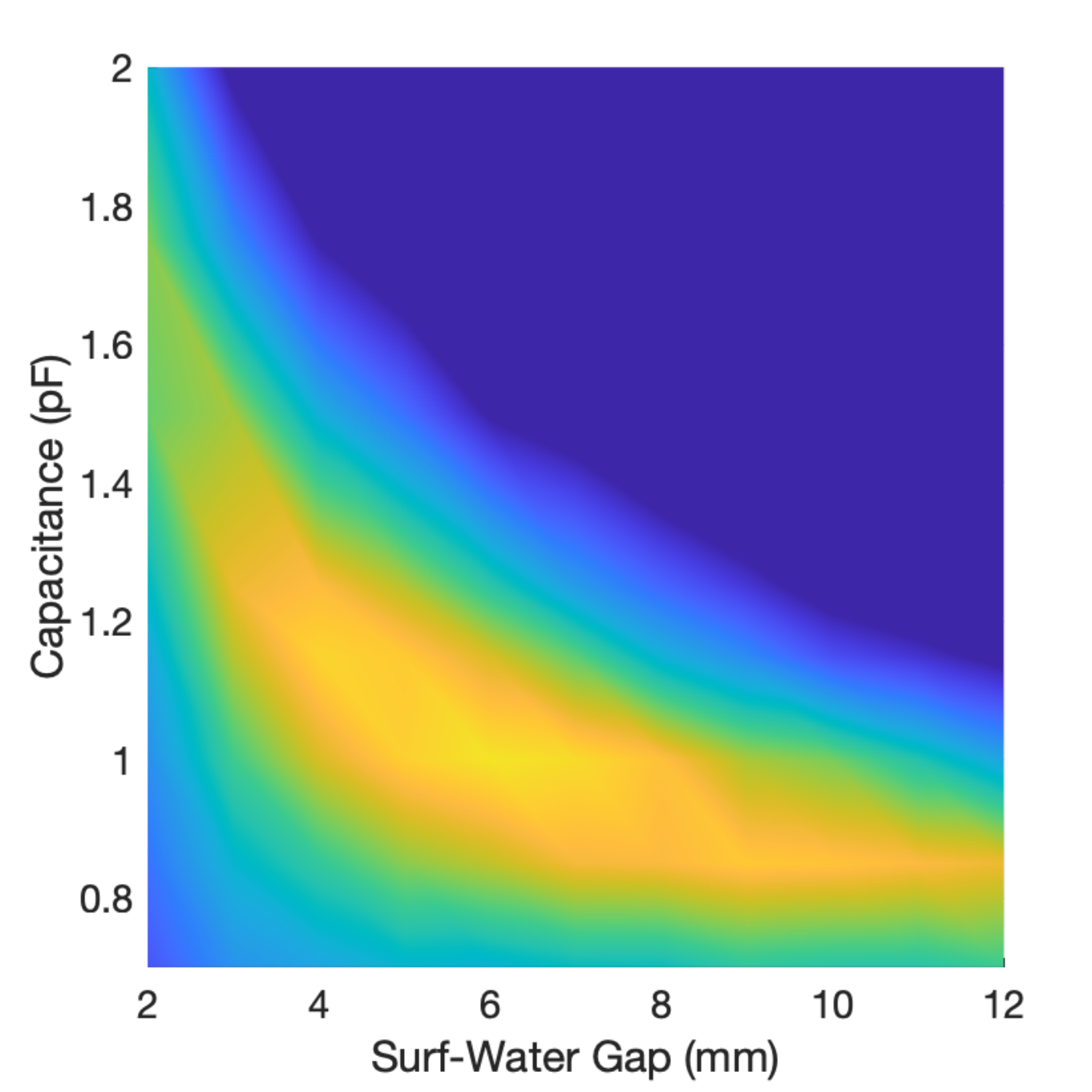}
  \caption{}
  \label{fig:design-hfss-water-gap-heatmap}
\end{subfigure}%
\begin{subfigure}{0.33\columnwidth}
  \centering
  \includegraphics[scale=0.2]{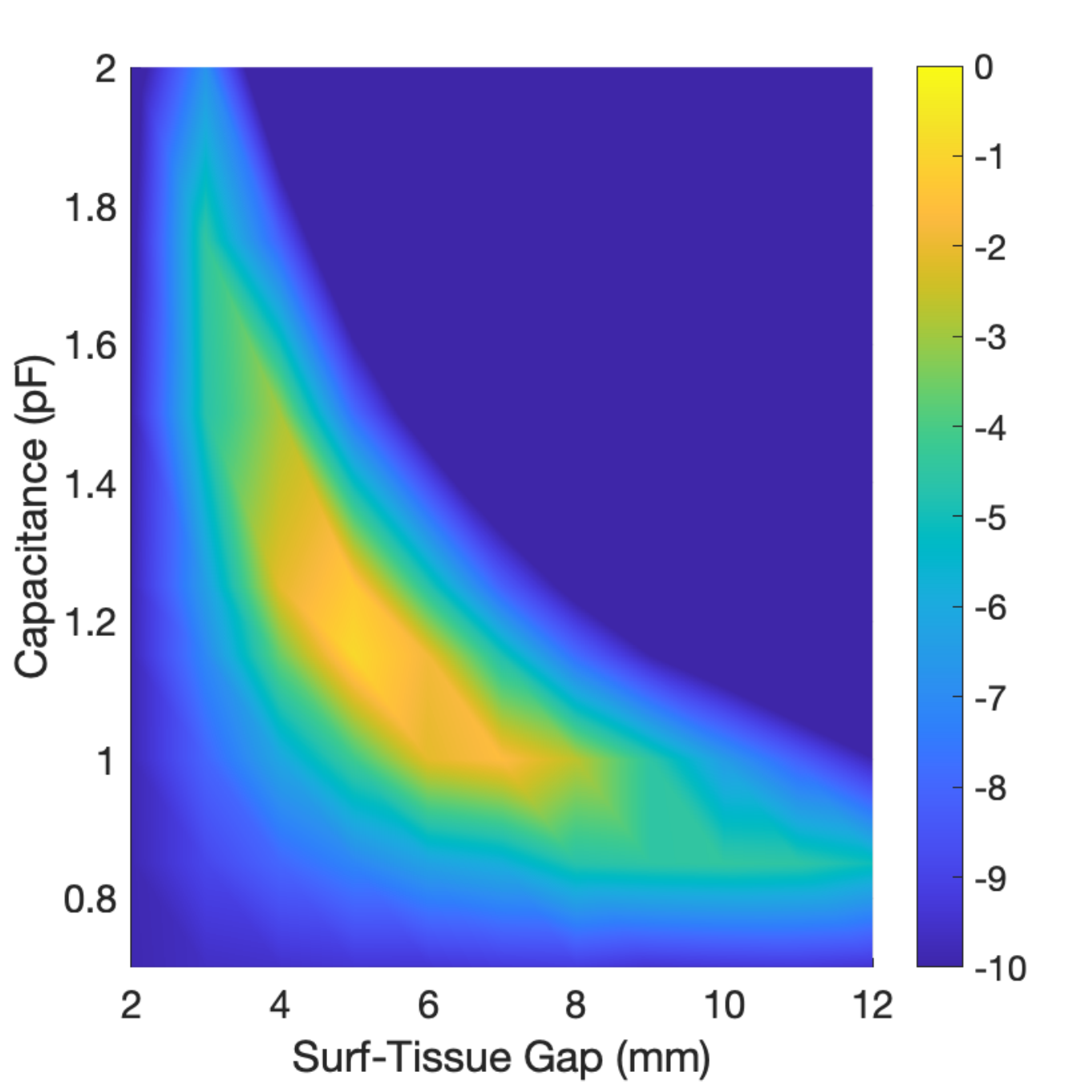}
  \caption{}
  \label{fig:design-hfss-tissue-gap-heatmap}
\end{subfigure}
\begin{subfigure}{0.33\columnwidth}
  \centering
  \includegraphics[scale=0.2]{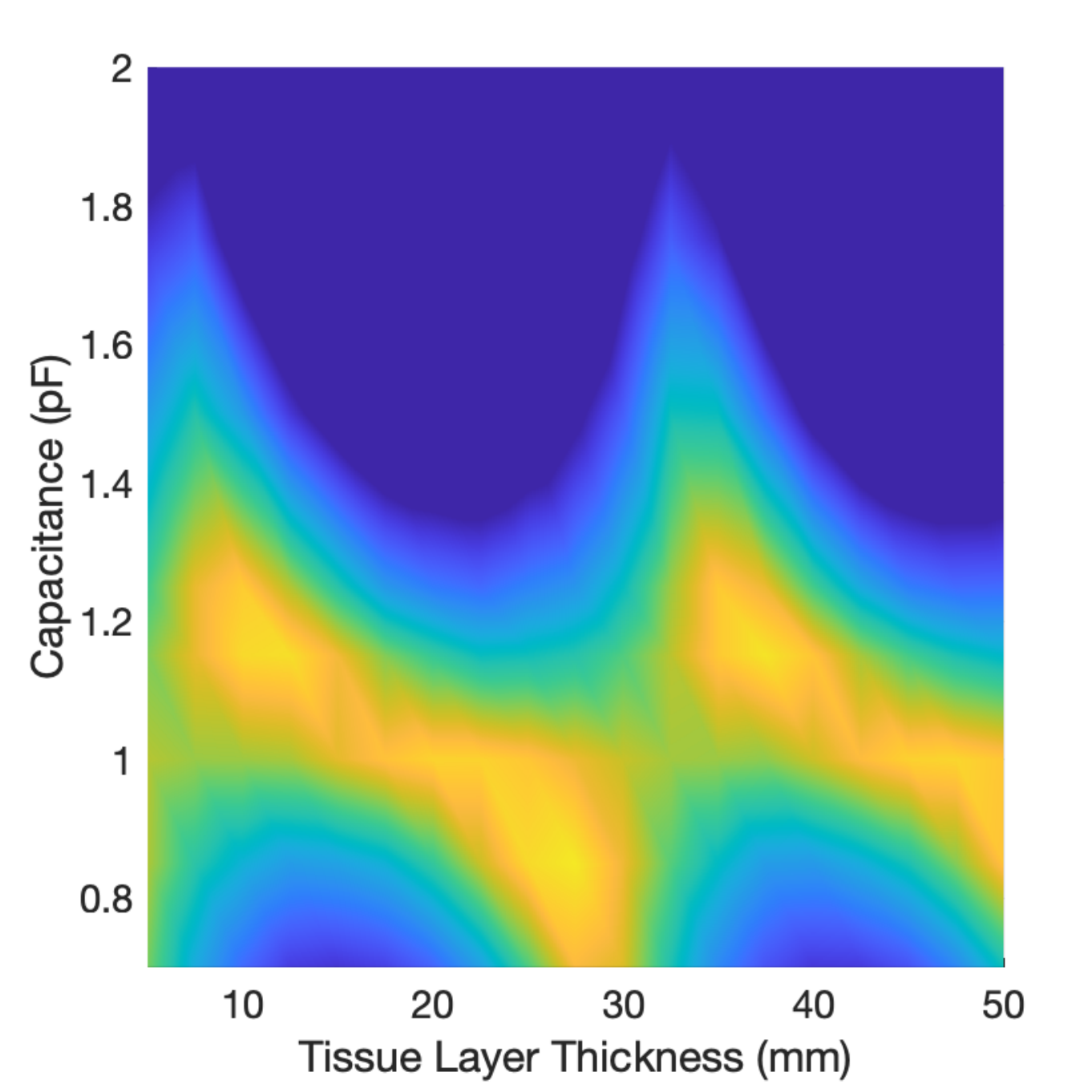}
  \caption{}
  \label{fig:design-hfss-tissue-fat-heatmap}
\end{subfigure}
\caption{Heatmaps of through-interface power vs. varactor capacitance. \textmd{By tuning the capacitance, we can maintain a high transmission power with different surface-media gaps \textbf{(a) (b)} or fat thickness \textbf{(c)} in HFSS.} }
\label{fig:design-hfss-heatmap}
\end{figure}

Previous simulations assume a fixed 6~mm surface-media gap and 15~mm fat thickness.
In \figref{fig:design-hfss-heatmap}, we verify the programmability provided by the varactors on the surface.
For surface-media gap from 2~mm to 12~mm and fat thickness from 5~mm to 50~mm, 
our surface design can maintain a high power transmitted through the interface by choosing a suitable capacitance value.
The distribution of appropriate capacitance values matches the distribution of admittance values in \figref{fig:design-theory-heatmap}, showing a good match with the theoretical model.

\heading{Beamforming gain.}
Endpoints in challenging media, like water or tissue, often experience destructive multipath fading causing significant power loss~\cite{remix-sigcomm18}.
As mentioned above, our surface has voltage control on each element.
Although we use such control for media matching, after finding the suitable capacitance, we still have 
control on whether each element should performs the matching.
So we have on-off control on whether signals pass through the surface, thus, our system can mitigate multipath fading and provide beamforming gain~\cite{rfocus}.
Fine grained phased control for beamforming is also possible with more layers of surface.
According to our simulations, overlaying 3 identical layers of our design can provide up to 180 degree phase control while maintaining a good media impedance match.
This can further improve beamforming performance, but we chose a 1-layer surface design for our system for its simplicity and light weight.

\heading{Surface size.}
Given the surface pattern for each element, we need to decide the size of surface, i.e., how many repetitive elements.
The previous simulations have the assumption of an infinite surface, while a finite surface can have different properties. We verify with simulations that a surface with $4\times4$ elements suffices. 
Above this number, the size of surface should match the application scenarios. 
A small surface may be favorable when the form factor is critical, for example, when integrated with garment as a wearable device; 
Otherwise, a larger surface can provide a larger coverage area and higher beamforming gain. 



\subsection{Control Algorithm}
Given the hardware design with element-wise control, 
we need an algorithm to search for the best config for each element efficiently.
The algorithm should be able to perform media impedance matching and beamforming simultaneously.

We consider the multipath channel between TX and RX
\begin{equation}
  h_{T-R} = h_{env} + \sum_{i=1}^{i=N} s(V_i)~h_{i}
\end{equation}
where $h_{env}$ is the channel of paths not going through the surface, $N$ is the number of surface elements, $s(V_i)$ is a complex number representing the amplitude and phase changes caused by surface element $i$, $V_i$ is the voltage applied, and $h_{i}$ is the channel of paths going through element $i$.
The algorithm goal is to set appropriate surface configuration, i.e., voltages for each surface element, to maximize received signal strength $\| h_{T_R}\|$ based on the endpoint feedback of signal strength.

\heading{Challenges.}
To achieve the goal,
We need to not only improve $\| s(V_i)\|$ to boost transmission, i.e., perform media impedance matching, 
but also combine signal paths constructively at the endpoint, i.e., avoiding multipath fading and providing beamforming gain.
For brute-force search, even if we consider 8 discrete voltage values only, for a surface with 64 elements, 
the search space ($8^{64}$) is too large for any practical deployment.
Thus, we aim for an efficient search algorithm.
The challenge, however, is that we can not separate media impedance matching effect from multipath effect at the endpoint feedback.
What's worse, the response of surface element $s(V_i)$ is not a fixed function of applied voltage $V_i$ that we can know in advance.
As shown in \figref{fig:design-hfss-heatmap}, other factors can easily affect the best capacitance(voltage) for media impedance matching.

\heading{Multi-stage solution.}
We divide and conquer this problem with a multi-stage control algorithm.
For the first stage, 
we uniformly apply voltage $V$ to all surface elements to probe $s(V)$.
The channel equation above can be written as 
\begin{equation}
    h_{T-R} = h_{env} + s(V) \sum_{i=1}^{i=N} h_{i}
\end{equation}
With a high probability, $\sum_{i=1}^{i=N} h_{i}$ is much larger than noise. 
The change of $s(V)$ can be observed from the change of $h_{T-R}$ when different voltages applied.
We record the two voltage states, $V_1$ and $V_0$, that maximizes and minimizes endpoint signal strength ($\|h_{T-R}\|$) as input for the next stage.
Note that without knowing how phase is aligned with $h_{env}$, we can not tell which state is maximizing $\| s(V_i)\|$ or which is minimizing.
Fortunately, we only need to know that these two states create opposite effects.

For the second stage, 
we use the afore-mentioned states ($V_1$ and $V_0$) as on-off control states and perform a randomized majority voting procedure to find the best on-off config.
Specifically, we want to determine the sets of elements taking on and off states, $S_1$ and $S_0$.
The channel we want to improve can be re-written as 
\begin{equation}
  h_{T-R} = h_{env} + s(V_0) \sum_{i\in S_0} h_{i} + s(V_1) \sum_{j\in S_1} h_{j}
\end{equation}
To do that, we generate random on-off configurations for the surface and record the received signal strength from endpoint feedback.
If a configuration leads to a higher signal strength than the median strength of all configurations, we cast a vote for each turned-on element in this config.
After testing all configurations, we count votes for each surface element.
We mark the elements as on ($V_1$) in the final config if it receives votes from more than half of the tested configs.
The output for this stage is the on-off config decided by such majority voting procedure.
For our surface with 64 element-wise control, we measure the feedback for 128 random configs as input, which is significantly fewer than all $2^{64}$ configs.
We find setting the number of random configs as twice of the number of control achieves a good balance between performance and feedback overhead.
The extra gain of testing more random configs is diminishing according to our experiments.
This stage is similar to the algorithm in RFocus~\cite{rfocus}, which is shown to be near-optimal for on-off control.
In \figref{fig:eval-algo-cdf-control}, we experimentally compare the performance of randomized procedure versus enumerating all configs.

For the third stage, we fine tune $V_1$ and $V_0$ voltages by testing a few adjacent voltages without changing the on-off config.
If a higher signal strength is received, we update $V_1$ and $V_0$ with the fine tuned voltage values.
This is to align the phase of $h_{env}$ and the phase of signal from surface and to account for any inaccuracy in stage one.

\heading{Control granularity.}
Although we have fine grained biasing voltage control for varactors, we only consider a set of 8 voltage values (30, 20, 15, 10, 5, 2.5, 0) in the control algorithm.
These values are chosen empirically to cover realizable surface admittance range. 
Considering more voltage values can bring additional performance gain, but requires more feedback measurements.
Another design choice about control granularity is that we utilize element-wise control of the surface.
In \figref{fig:eval-algo-cdf-control}, we demonstrate the extra gain from element-wise control.

\heading{Feedback mechanisum.} 
We reply on the feedback from endpoints to guide our system since we don't incorporate hardware for sensing as in LAVA~\cite{lava-sigcomm21}.
Specifically, we use the received signal strength (RSS) at the endpoints as feedback, which is accessible across different wireless hardware.
Due to the reciprocity of metasurface operation, feedback from either endpoint is sufficient.
For backscatters, we use the received backscatter signal strength measurement from the endpoint in air, similar to In-N-out~\cite{inNout-mobicom20}.






\subsection{Discussion}
\label{sec:design-diss}

We discuss other applications that can be enabled by tuning media interfaces.

\heading{Supporting in-air links.}
Since our system controls the reflection from the media interface,
it can provide link enhancement for in-air links similar to existing surface systems in air.
For example, a user wearing the surface for in-vivo networking can also get power boost for near-by devices such as phones or smart watches.
We evaluate such link support in \secref{sec:eval-link-air}.

\heading{Protection against wireless tracking.}
Recent work~\cite{rf-protect, wi-peep-mobicom22} look into the security issues of wireless sensing.
Through-wall wirelss sensing can challenge the perception of indoor privacy.
We note that wirelss sensing relies on the reflection from human body.
Our surface can significantly reduce the reflection, like invisible clocks, or dynamically program the reflection to protect users against unwanted sensing systems.

\heading{Media sensing.}
Our system adapts to the changing environment and different media with programmability.
We can consider the reverse process for sensing environmental information.
Specifically, by observing the voltage tha we achieve media impedance matching, 
we can sense information about surrounding media. 
This is out of the scope of this work.

%% file: implementation.tex
\begin{figure}[t]
  \begin{subfigure}{0.9\columnwidth}
    \centering
    \includegraphics[width=0.95\columnwidth]{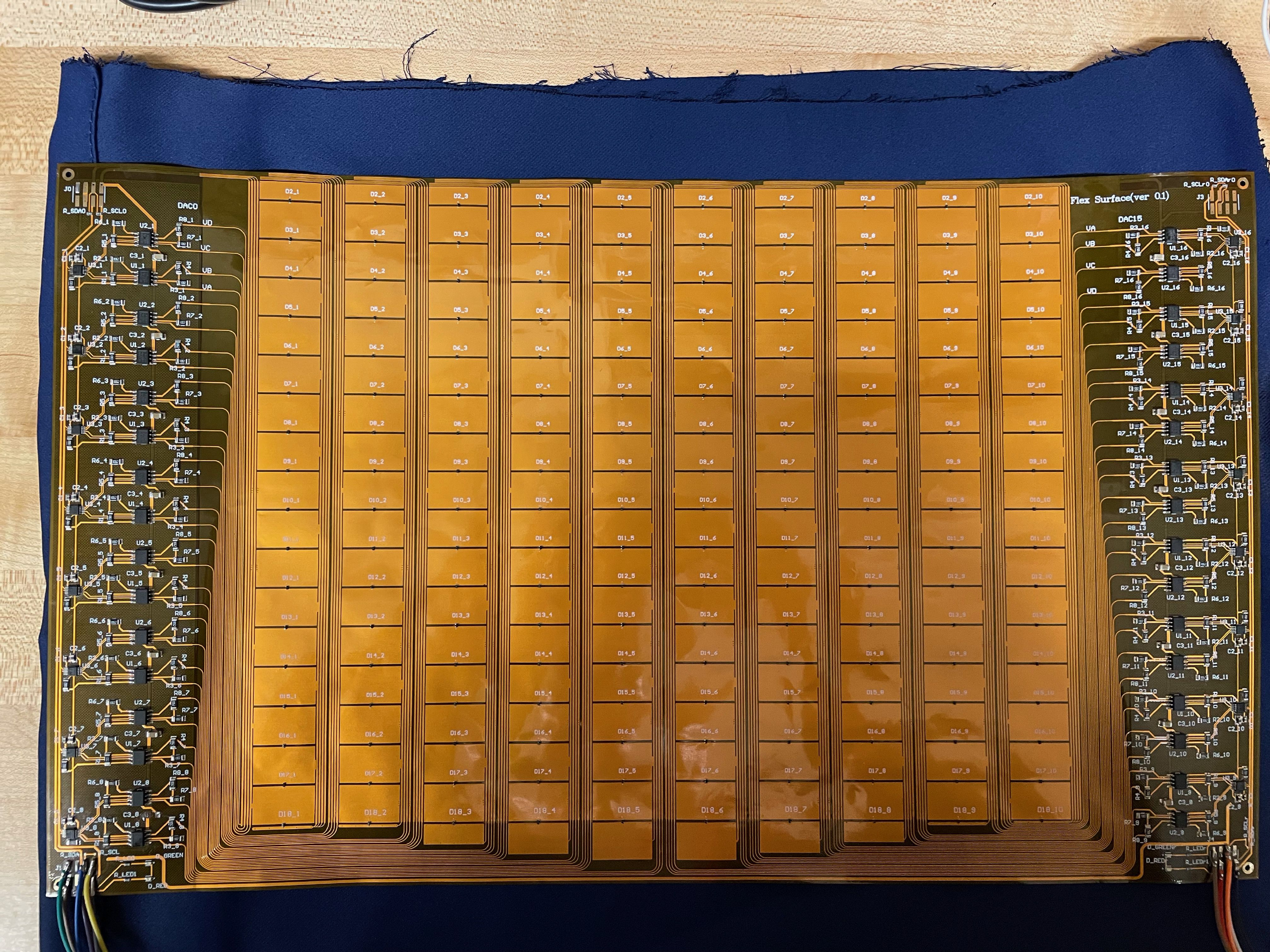}
    \caption{Surface prototype as a flexible PCB. 
    \textmd{The control circuit is currently placed on the sides to ease soldering. 
 Further circuit integration can enable a smaller form factor.} 
    }
    \label{fig:impl-surf-front}
  \end{subfigure}
  \bigskip\\
  \begin{subfigure}{0.45\columnwidth}
    \centering
    \includegraphics[width=0.95\columnwidth]{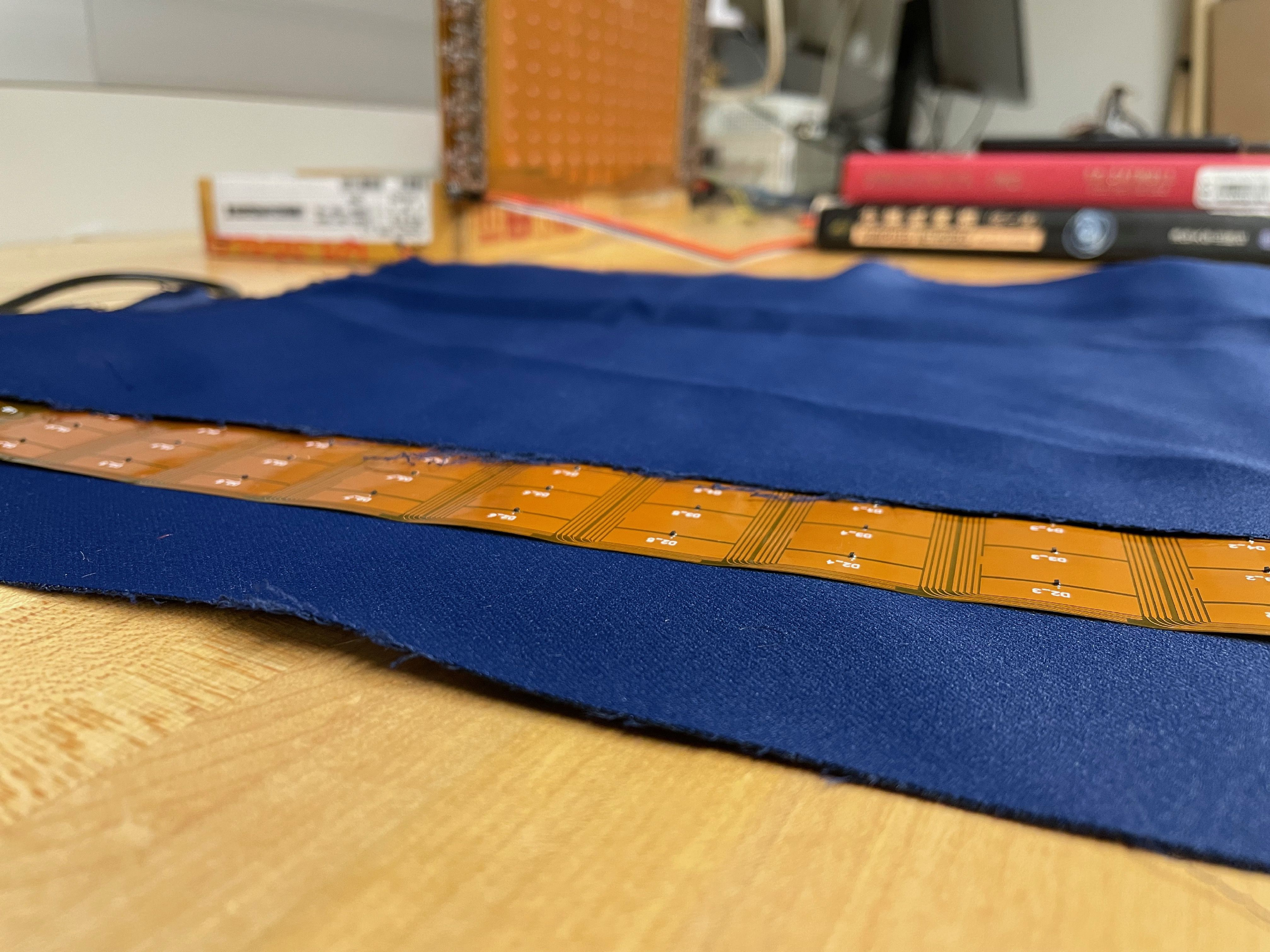}
    \caption{0.1~mm thickness.}
    \label{fig:impl-surf-thin}
  \end{subfigure}%
  \begin{subfigure}{0.45\columnwidth}
    \centering
    \includegraphics[width=0.95\columnwidth]{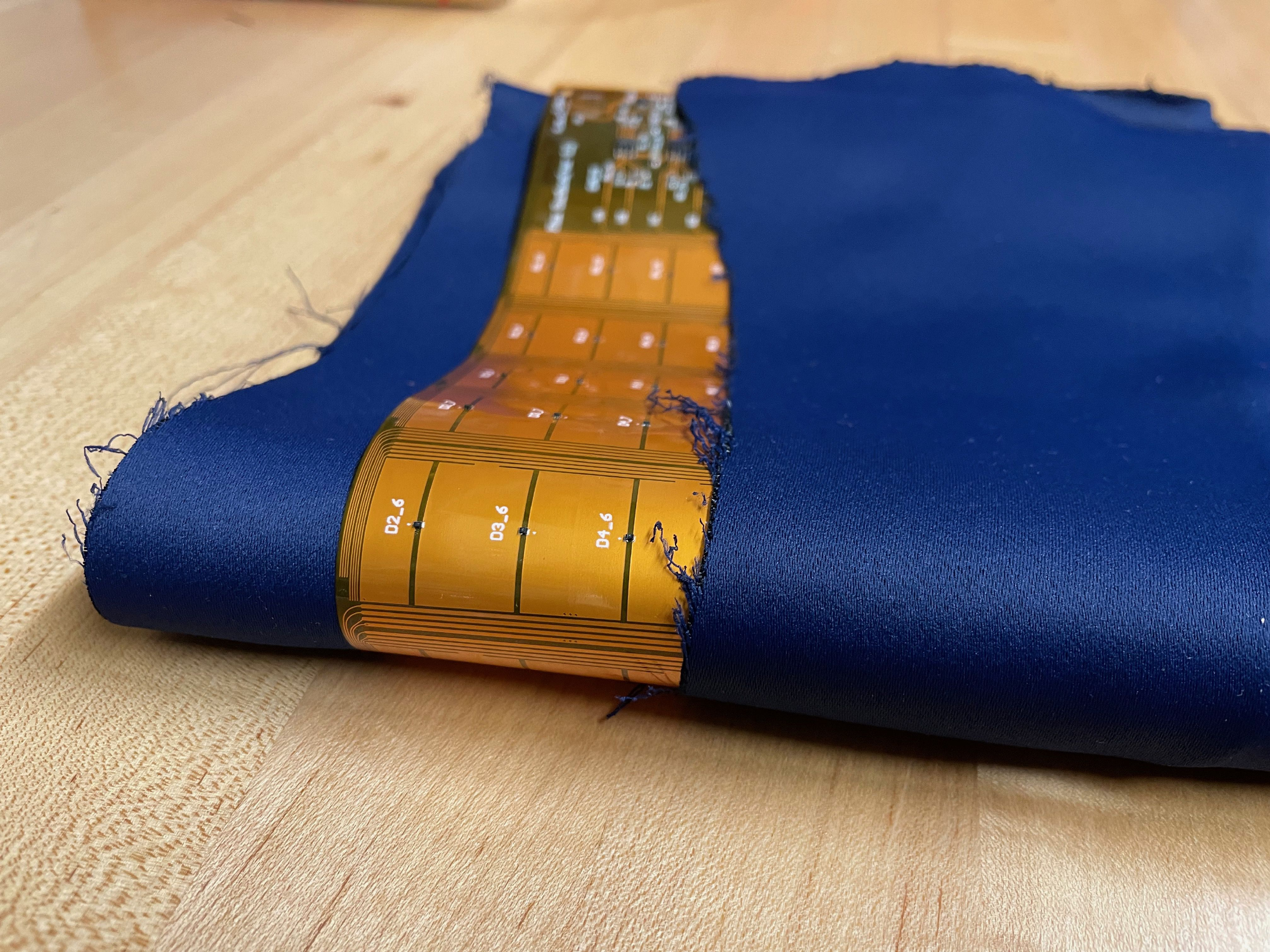}
    \caption{As flexible as fabric.}
    \label{fig:impl-surf-flex}
  \end{subfigure}%
  \caption{Surface hardware implementation.}
  \label{fig:impl-surfaces}
\end{figure}

\section{Implementation}
\label{sec:impl}

\heading{Surface Fabrication.}
We fabricate the surface using regular commercial flexible PCB fabrication process and manually solder circuit components including the SMV1405 varactors~\cite{SMV1405} (\figref{fig:impl-surf-front}).
A surface prototype is as thin as a piece of paper, about 0.1~mm thick, (\figref{fig:impl-surf-thin}) and as flexible as fabric (\figref{fig:impl-surf-flex}).
This facilitates flexible surface deployment, 
e.g., as a wearable metasurface for medical monitoring. 
Due to limitations of the fabrication process, the size of our surface is capped at 24$\times$40~cm, with $8\times10$ elements. 
We individually control the voltage of 64 elements, while the remaining 16 act as padding to avoid surface edge effects.
Per surface, the PCB fabrication cost \$50 and the circuit components around \$100, or \$1.5 per tunable element.

\heading{Control Circuit.}
We exploit the fact that varactors hardly draw any current and design a lightweight voltage control circuit, using DAC chips to output voltages with fine-grained control and amplify the voltages with operational amplifiers to the desirable voltage range.
This contrasts with previous programmable metasurfaces~\cite{llama,rflens_2021} that relied on bulky external circuits or devices for varactor voltage control, which could hinder their practical deployment.
For our implementation, we use 16 4-channel DAC MPC4728 chips~\cite{mpc4728} and connect each chip to 2 LM358 op amplifiers~\cite{lm358} to amplify the voltage from 0--4~V to 0--36~V.
We prioritize the ease of manual soldering for the current implementation, while a highly integrated circuit implementation can enable a smaller form factor.
An ESP32 micro-controller~\cite{esp32-devkitc} is used to communicate with DAC chips through 2 I2C buses.
We run the control algorithm with Matlab on a PC with Intel I7-7700 and control the ESP32 over Wi-Fi.

\heading{Control speed.}
The speed of control dictates whether our system can react promptly to fluctuations in the perceived propagation medium characteristics and multipath fading behavior within each medium. 
With the high speed mode, each I2C bus in our control circuit provides a bandwidth of 3.4~Mbps~\cite{mpc4728}.
It takes 9 bytes to transfer the voltage data to one chip and only 0.17~ms to set 16 chips with 2 buses.
Our control algorithm goes through 3 exploration stages, which requires 8, 128, and 9 tests of surface configurations respectively.
Putting them together, our system can explore the search space in less than 25~ms at max speed. 
This is well within the channel coherence time for most scenarios and enables mobility support with real-time adaption, although we do not explicitly optimize for endpoint mobility in the current implementation.


\heading{Power consumption.}
Our surface design is passive and consumes little power. 
The varactors are reverse-biased and draw a maximum of 20~nA of reversed current, or 
less than 1.2~uW power per surface element under a 30~V bias voltage.

%% file: eval.tex
\section{Evaluation}
\label{sec:eval}

\begin{figure}[t]
  \centering
\begin{subfigure}{0.45\columnwidth}
  \centering
  \includegraphics[scale=0.13]{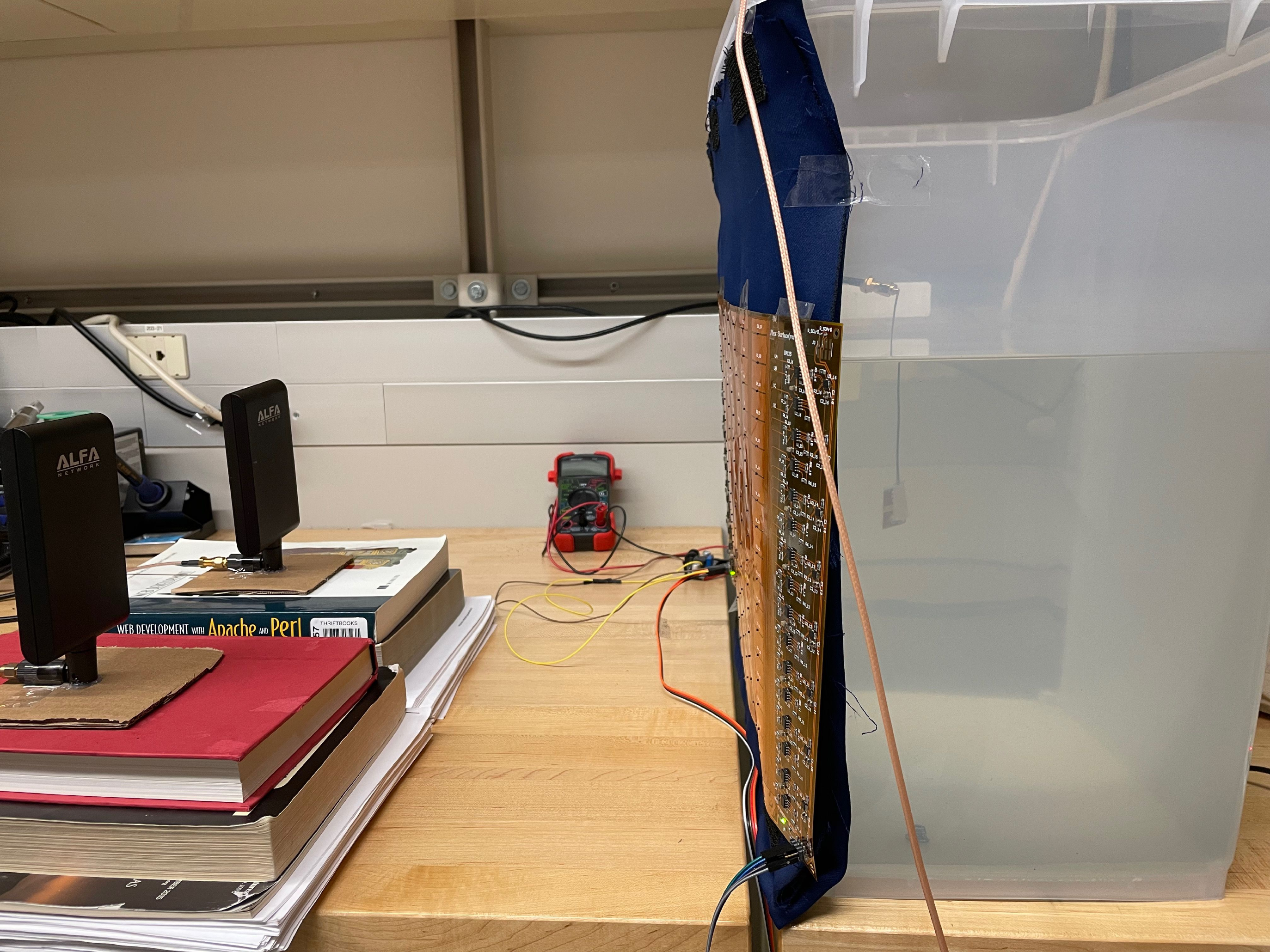}
  \caption{}
  \label{fig:design-setup-water}
\end{subfigure}%
\begin{subfigure}{0.45\columnwidth}
  \centering
  \includegraphics[scale=0.13]{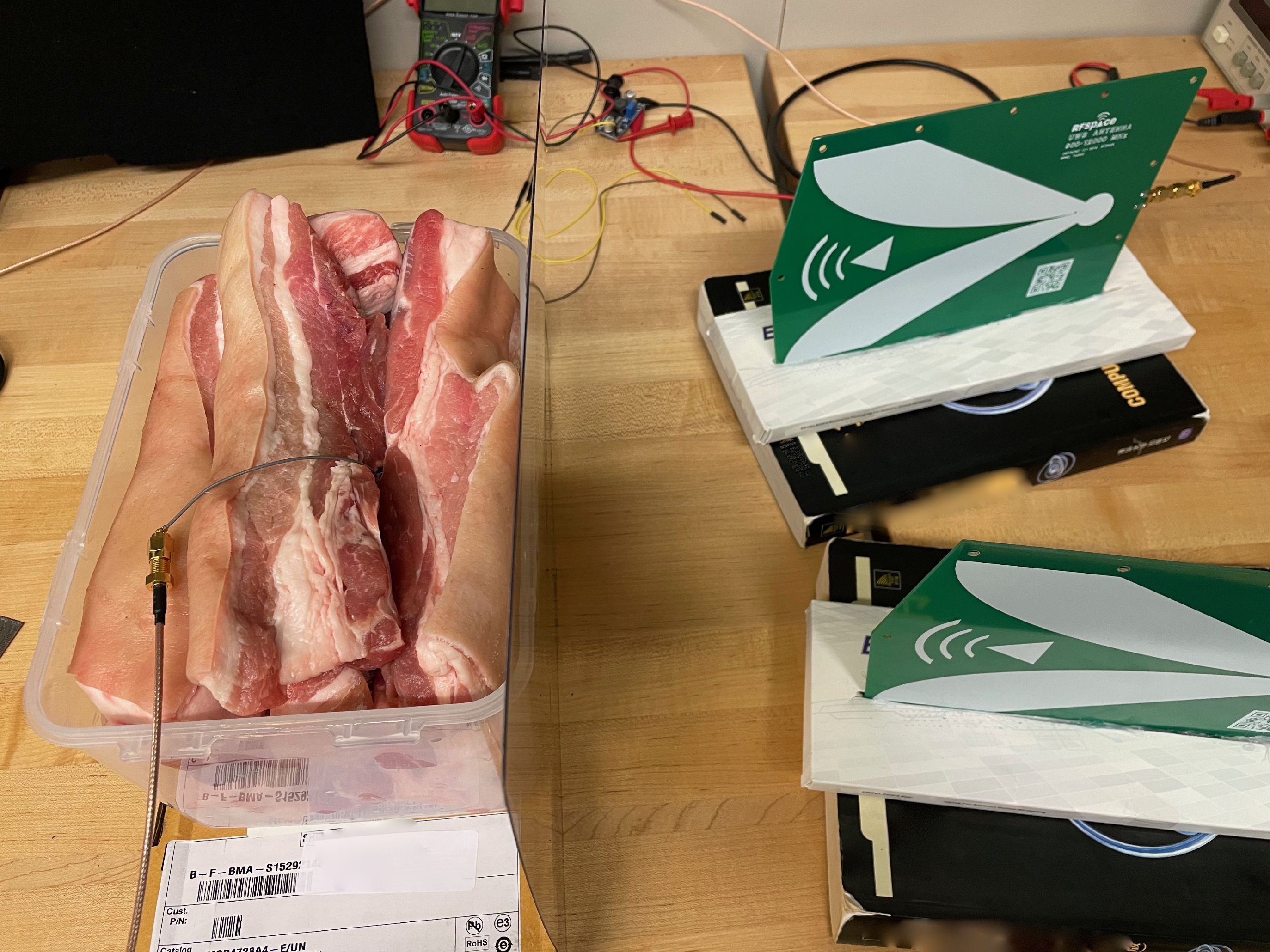}
  \caption{}
  \label{fig:eval-setup-tissue}
\end{subfigure}
\caption{Experiment setups. \textmd{\textbf{(a)} Water bucket with the surface and fabrics. \textbf{(b)} A box of pork belly with an antenna inside. \todoEdit{Both setups have gaps between surface and media.}}}
\label{fig:eval-setups}
\end{figure}

\heading{Media studied.} 
We mainly study the performance of air-water and air-tissue (emulated with pork belly) links. 
The property of ground (soil) is dictated by the water it contains~\cite{soil_permittivity_factors}, so we expect the performance for air-to-ground links to be similar to air-to-water links.
We place water in a bucket and pork belly in a container (\figref{fig:eval-setups}).
Although there is a layer of plastic between air and the other test medium, 
the permittivity of plastic is very close to air compared to water and tissue, thus, the plastic can be treat as air with respect to signal propagation behavior.
The surface is placed in front of the bucket and container with a default gap of 6mm.
We control the gap size with the number of fabric layers or air in between.

\begin{figure}[t]
  \centering
\begin{subfigure}{0.45\columnwidth}
  \centering
  \includegraphics[scale=0.3]{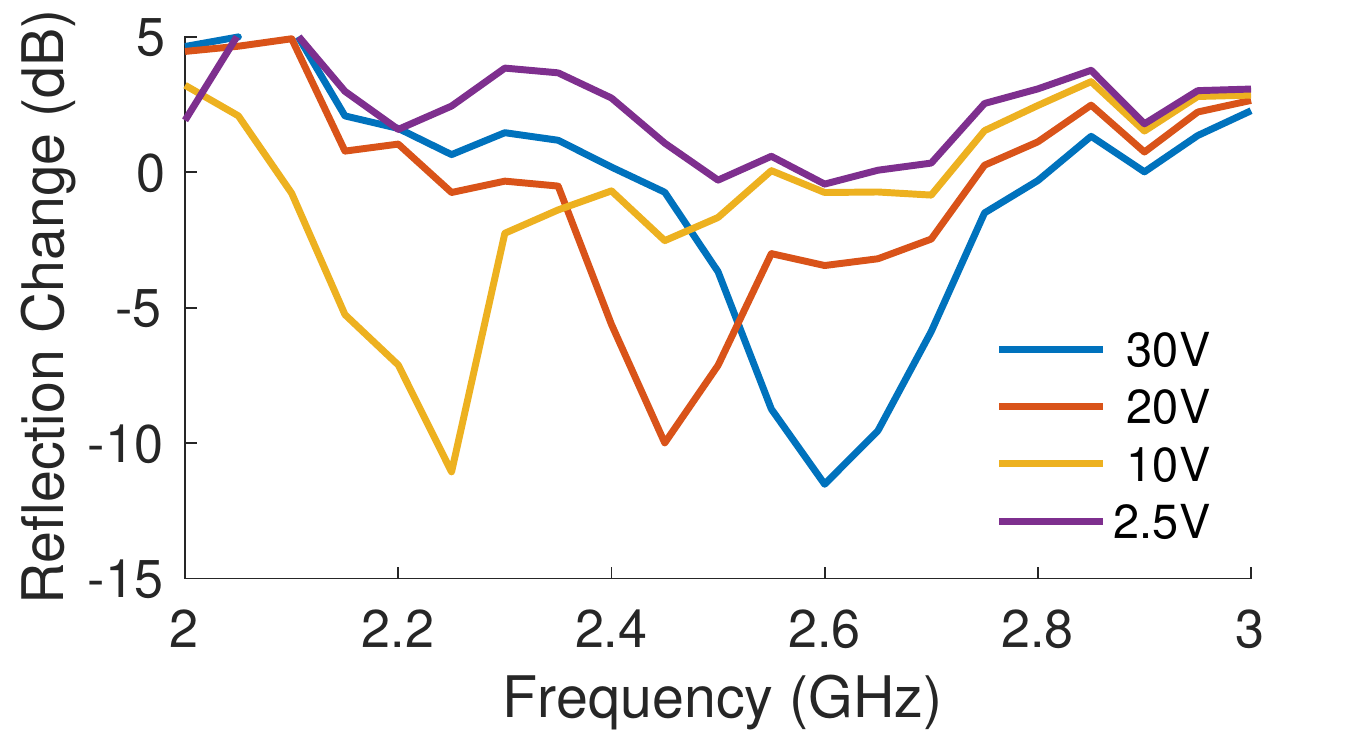}
  \caption{}
  \label{fig:design-reflex-water}
\end{subfigure}%
\begin{subfigure}{0.45\columnwidth}
  \centering
  \includegraphics[scale=0.3]{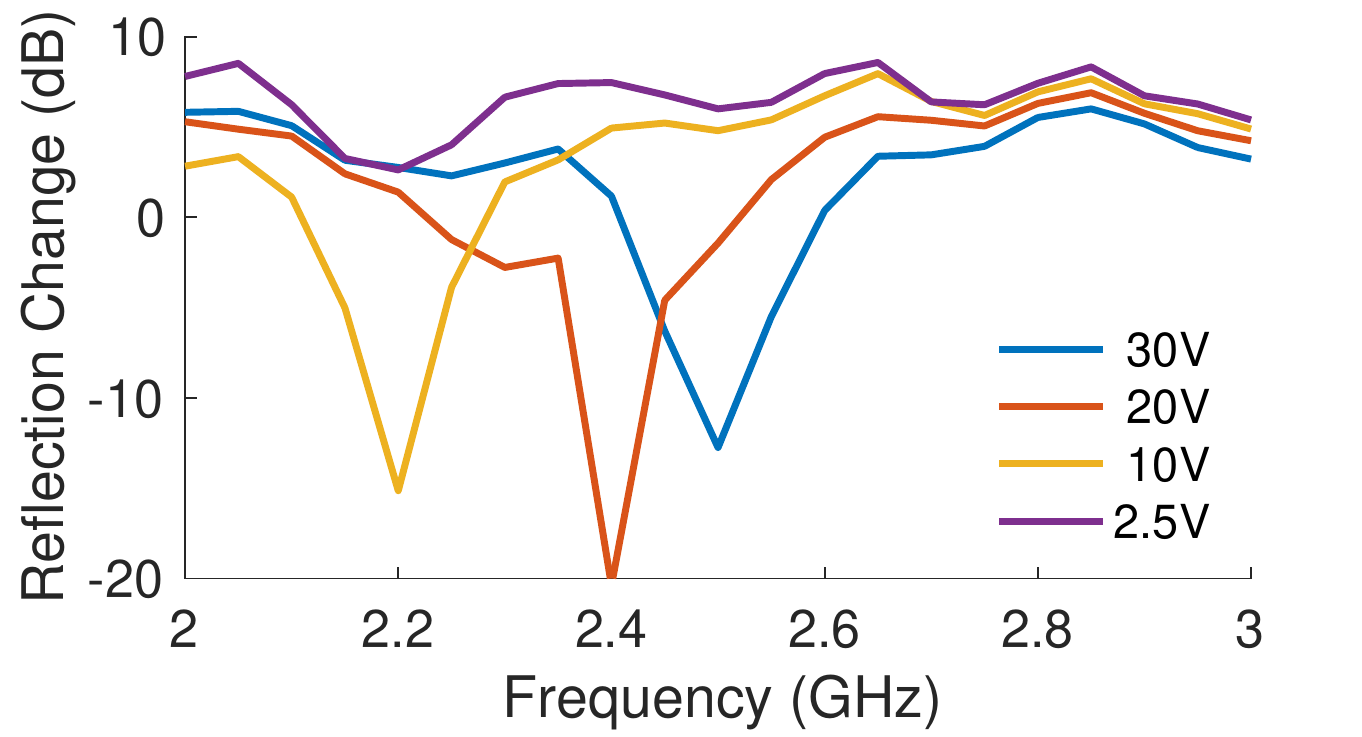}
  \caption{}
  \label{fig:eval-reflex-tissue}
\end{subfigure}
\caption{Reflection reduction over frequency. \textmd{Matching media impedance reduces reflection by over 10~dB for water~\textbf{(a)} and tissue~\textbf{(b)}. The biasing voltage should be set to match the center frequency of the links (the red line, trough showing minimal reflection).}}
\label{fig:eval-reflex}
\end{figure}

We conduct most experiments with the water bucket, which is similar in size to human chest.
It is not only to show the enhancement for air-to-water links, but also to demonstrate gain for air-to-tissue links.
Impedance matching for water and tissue follows the same principle described in \secref{sec:design-principle}, and in-water devices experience similarly difficult channel conditions as in-body devices~\cite{ivn-sigcomm18}.
It is much easier to experiment with a large amount of water and sample locations inside it than tissue.
We show the system performance for pork belly in \figref{fig:eval-reflex-tissue} and \figref{fig:eval-water-tissue-reciprocity-gain}.

\heading{Link Setup.}
We set up 2.4~GHz links for experiments with two USRP N210s~\cite{usrp-n210} unless otherwise noted.
We use directional patch antennas~\cite{patch_antenna} for endpoints in the air and use omni-directional flexible antennas~\cite{flexnotch-antenna} for endpoints in the water and tissue.
We place the (water-proof) flexible antennas inside water or tissue and connect to USRP with a coaxial cable.
To sample different channel conditions, we move both endpoints around and create different incident angles for the incoming signal to surface, up to an angle of 45 degrees.
We use the received signal strength of in-air endpoint as feedback for all experiments, since it is easier to acquire than feedback from any in-water/tissue endpoint.
According to our measurements, due to channel reciprocity, the channel feedback in both directions only differ by a constant, so both can guide the surface control algorithm and provide similar results.
For all links in the following experiments, we remove the surface as a faithful baseline to calculate the performance gain.

\begin{figure}[t]
  \centering
\begin{subfigure}{0.45\columnwidth}
  \centering
  \includegraphics[scale=0.3]{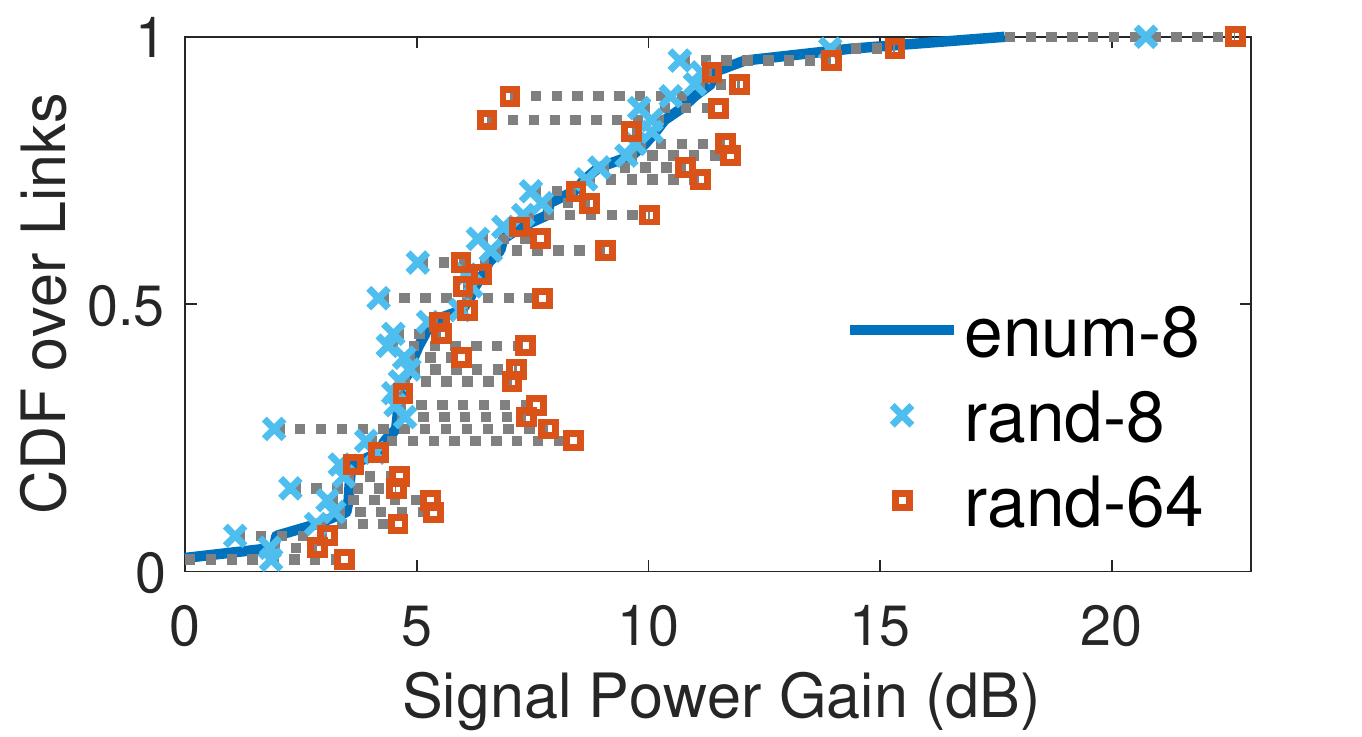}
  \caption{}
  \label{fig:eval-algo-cdf-control}
\end{subfigure}%
\begin{subfigure}{0.45\columnwidth}
  \centering
  \includegraphics[scale=0.3]{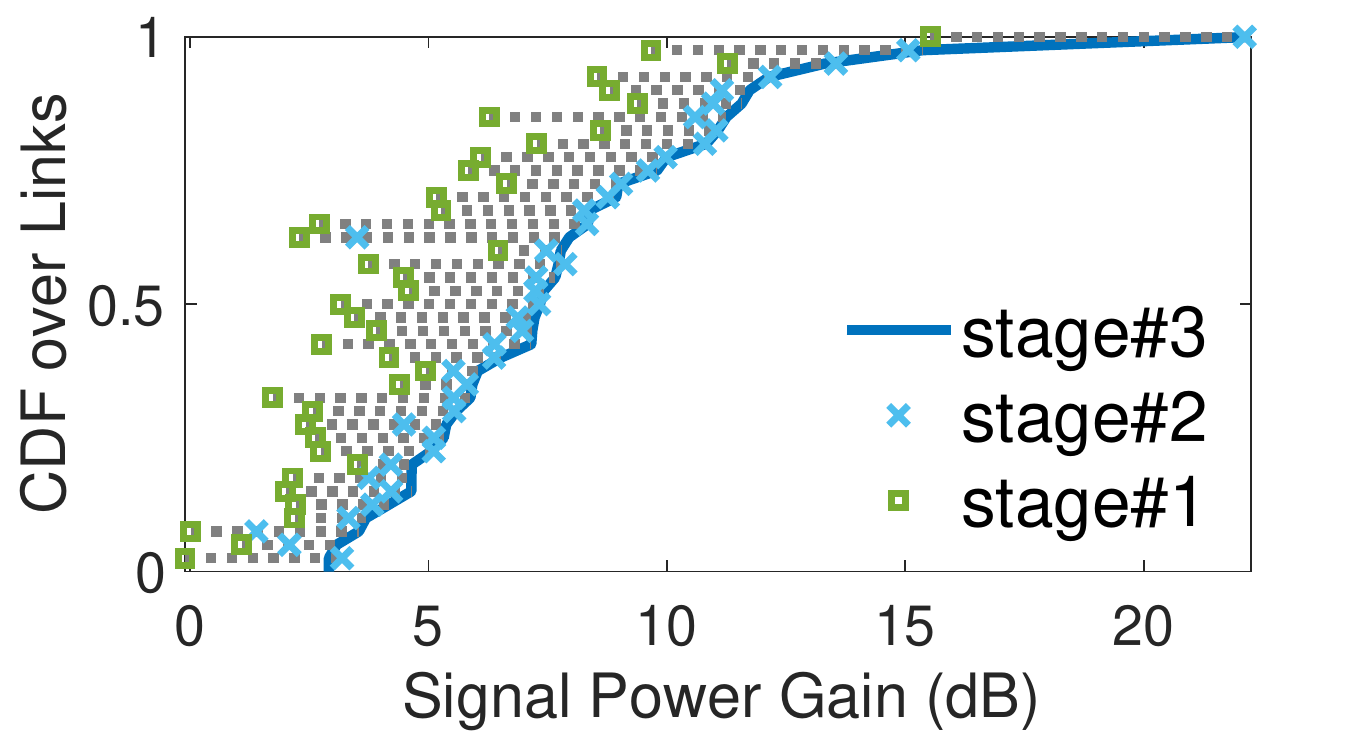}
  \caption{}
  \label{fig:eval-algo-cdf-stages}
\end{subfigure}
\caption{Control algorithm microbenchmarks. \textmd{\textbf{(a)} Randomized algorithm provides near-optimal performance; utilizing individual control of all 64 elements provides more gain. \textbf{(b)} Gain of 3 algorithm stages.}}
\label{fig:eval-algo-cdf}
\end{figure}

\subsection{Media Matching}
We first verify the fundamental functionality of the surface, i.e., impedance matching performance of two media.
Although our system can handle multipath effects with the control algorithm,
we set uniform voltage to all surface elements here to focus on the media matching part.
To avoid the interference from multipath, we place RF absorbing foams~\cite{absorbing-foams} around the experiment setup.
However, the received signal of in-water or in-vivo endpoint is still susceptible to multipath effect inside the water/tissue containers and hard to measure.
Thus, we measure the reflection power from the water and tissue, which is stronger and easier to measure.
The reduction of reflection should lead to increased signal power transmitted through the media interface.

We set up a link with two directional vivaldi antennas pointing towards the surface and media at an angle, as shown in \figref{fig:eval-setup-tissue} .
The antennas are placed around 30~cm in front of the water bucket and tissue container. 
This distance roughly ensures we only measure the most direct reflection path between the two antennas, bounced from the surface and media interface.
We first measure the received reflection strength with no surface deployed as the baseline.
Then, we measure the reflection strength when the surface is deployed and different bias voltages are applied.
\figref{fig:eval-reflex} shows the surface reduces the reflection by over 10~dB for both water and tissue around 2.4~GHz.
Appropriate voltages can achieve a good media impedance matching, i.e., large reflection reduction.
The shift of matching troughs with decreasing voltage agrees with the simulation results in \figref{fig:design-hfss-reflex-over-freq},
which indicates that the surface admittance is increasing as expected when the voltage decreases.
The exact voltage needed for media matching on 2.4~GHz is different from the simulated value, i.e., 20V instead of 10V.
This is because practical deployments can hardly match simulation perfectly, due to surface-medium gaps and tissue fat thickness.
This highlights the necessity of 
programmability for dynamic tuning in our design.

\subsection{Boosting Cross-Media Links}
Next, we evaluate performance of \name{} when boosting cross-media links with the control algorithm.


\heading{Control algorithm microbenchmarks.}
We benchmark the algorithm in terms of its performance with different configurations and performance at each stage.
For each setting, we set up 45 air-to-water links and measure the gain for received signal power at the in-water endpoint.
Specifically, we place the in-water antenna at 3 depths (5, 7.5, 10~cm), 5 different locations at each depth, while the in-air endpoint is placed at 3 locations (25, 35, 50cm away from the bucket).
We repeat the experiments with 3 algorithm variants, i.e., enumerating all states of 8 columns, column-wise, with $2^8$ tests, randomized voting for 8 columns with 32 tests, and randomized voting for 64 elements with 128 tests.
The results are shown in \figref{fig:eval-algo-cdf-control} using enumerating 8 columns as the reference.
Randomized voting for 8 columns shows near-optimal performance compared to enumerating 8 columns, but using much fewer tests.
Randomized voting for 64 elements provides higher gains for most links compared to 8 columns, 
showing the benefit of fine-grained element-wise control.
For the following experiments, we use randomized voting for 64 elements with 128 tests since it performs best.

Next, we characterize the gain we get from each stage of the algorithm with 40 links (\figref{fig:eval-algo-cdf-stages}).
All stages are necessary to provide gain, while the first two stages provide most of the gain.
How much gain we can achieve for a link is influenced by the specific multipath channel condition.

\begin{figure*}[t]
  \begin{minipage}{.32\textwidth}
    \centering
    \includegraphics[width=0.95\columnwidth]{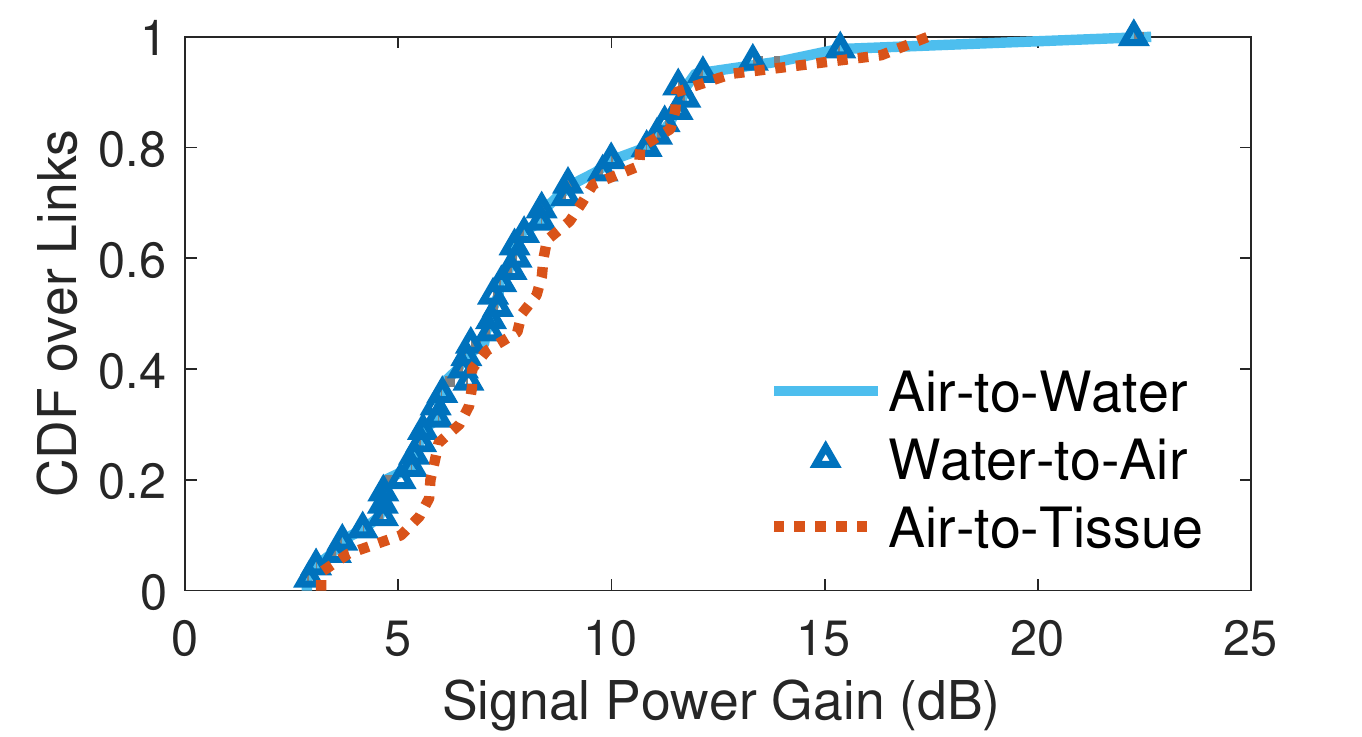}
    \caption{Received signal power gain for cross-media links.}
    \label{fig:eval-water-tissue-reciprocity-gain}
  \end{minipage}\hfill
  \begin{minipage}{.32\textwidth}
    \centering
    \includegraphics[width=0.95\columnwidth]{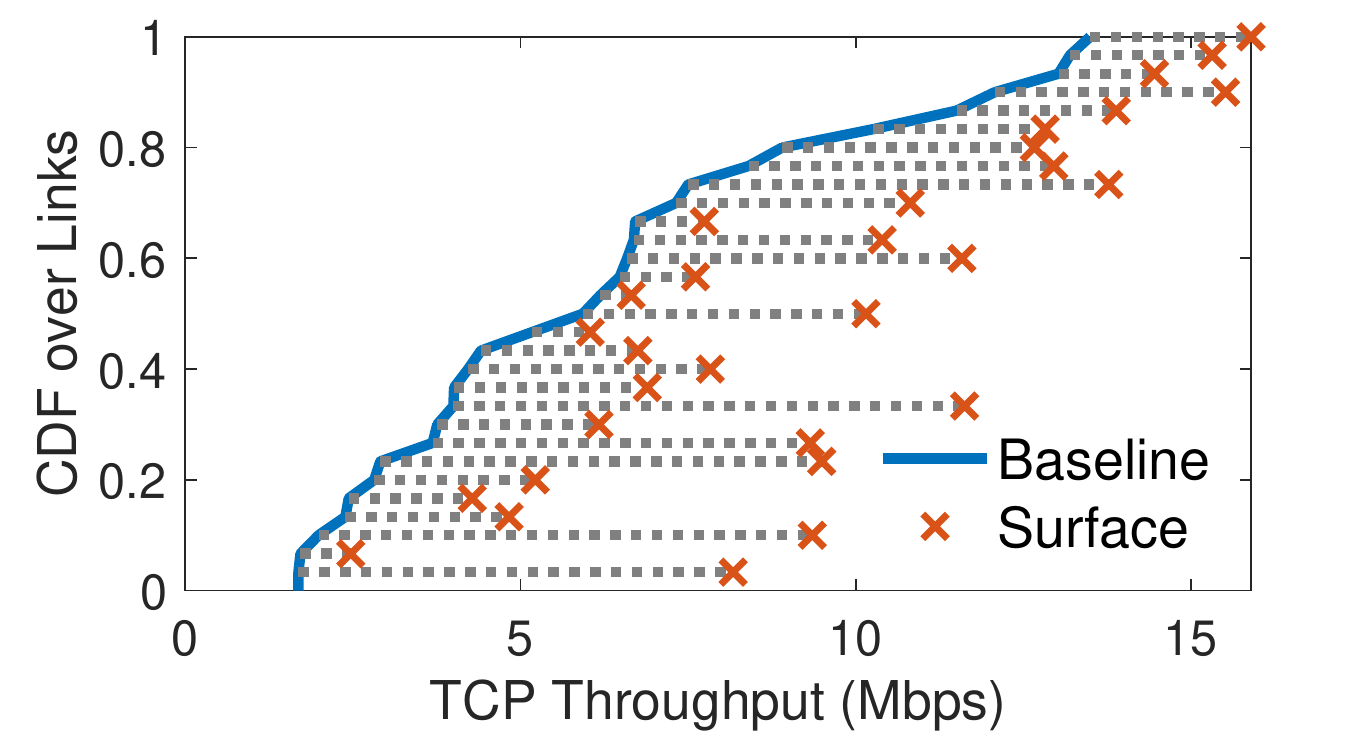}
    \caption{TCP throughput improvement for air-water Wi-Fi links.}
    \label{fig:eval-water-tcp-thru}
  \end{minipage}\hfill
  \begin{minipage}{.32\textwidth}
    \centering
    \includegraphics[width=0.95\columnwidth]{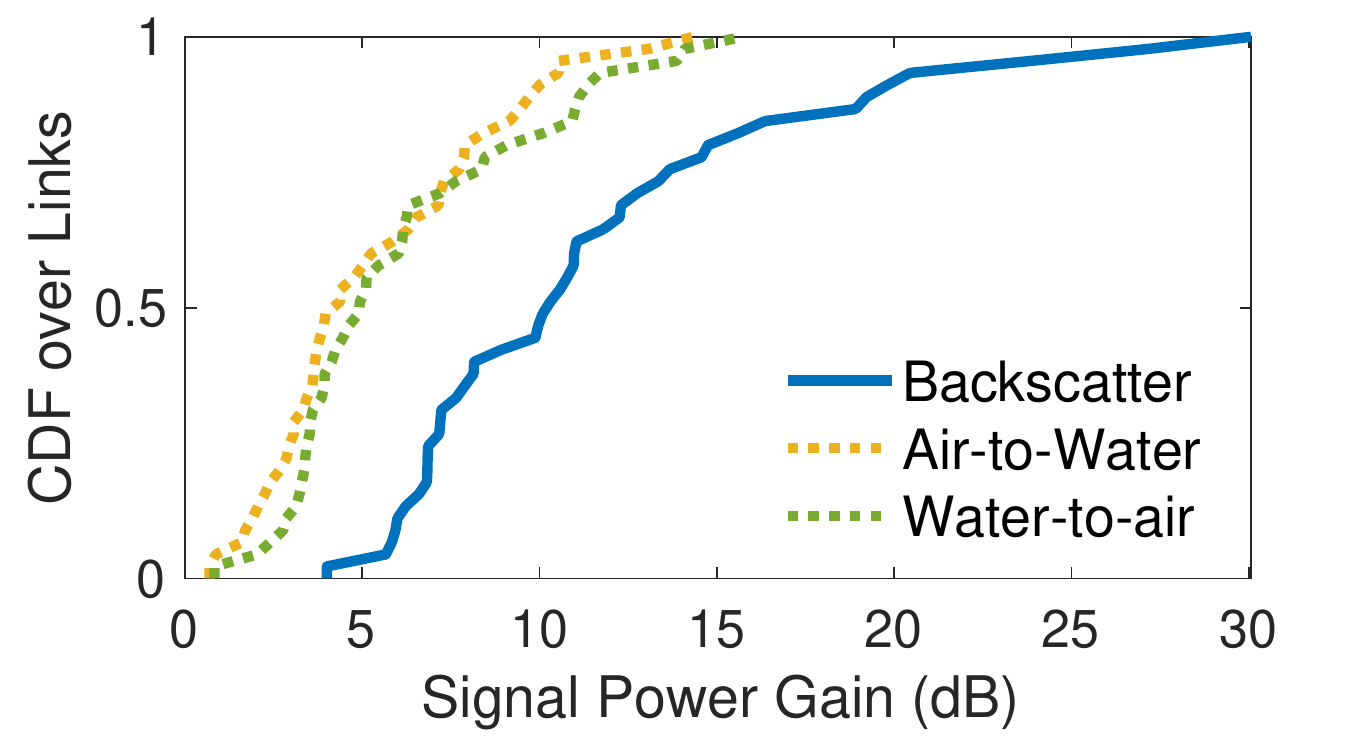}
    \caption{Gain for emulated backscatter links.}
    \label{fig:eval-backscatter-gain}
  \end{minipage}
\end{figure*}

\heading{Gain for cross-media links.}
We now evaluate the gain for air-to-water and air-to-tissue links as shown in \figref{fig:eval-water-tissue-reciprocity-gain}.
We set up 45 air-to-water links with the same setup mentioned above.
Our system boosts the received signal power of in-water endpoint by a median of 7.2~dB and up to 22~dB gain.
We also measure power gain for the reversed link direction, i.e., the received signal power gain for the in-air endpoint.
The gains in two directions match well, confirming that the effect of our surface is the same for both directions due to channel reciprocity.
For air-to-tissue links, we measure 30 links, with 2 in-vivo locations, and 15 in-air endpoint locations.
The results show a median gain of 8.2~dB and up to 17.5~dB for in-vivo received signal power.
This shows that our system can effectively match the impedance between the air and tissue despite the challenging multi-layer structure of tissue.
The median gain for tissue is higher than the median gain for water.
This agrees with simulation results in \figref{fig:design-hfss-backscatter-gain}, but the extra gain is smaller.
There are several possible reasons.
We use multiple pieces of pork belly to emulate the in-vivo environment, but there is still air gap in between.
This can lead to a higher baseline power than real in-vivo and simulated environments, thus lowering the gain.
Another reason is that each piece of pork has different fat thickness, creating large discontinuity of medium composition that would not normally happen, thus making media matching more challenging.


\heading{Throughput improvement.}
To show the throughput increase brought by the signal power gain, 
we measure TCP throughput with iperf for 45 air-to-water Wi-Fi links w/ and w/o the surface.
We use widely used IoT devices, ESP32~\cite{esp32-devkitc}, as the endpoints.
The control algorithm uses Wi-Fi RSSI from the in-air endpoint as the feedback.
We place a flexible antenna~\cite{flexnotch-antenna} in the water and connect it to ESP32 with a cable as the in-water endpoint.
TX power is set to 0~dBm to match the setting for devices with a tight power budget.
\figref{fig:eval-water-tcp-thru} shows the throughput with and without the surface. Our system provides throughput improvement to all links.
The median throughput increase is 55\%, while the maximum is around 400\%.

\heading{Backscatter links.}
For backscatter links, we use 2 USRP N210s as the in-air transmitter and receiver, and another USRP connected to a flexible antenna in water as the backscatter device.
This allows us to measure transmitter-to-backscatter and backscatter-to-receiver channels separately.
Since backscatter signals traverse both channels until reaching the receiver,
we multiply the channel coefficients in both directions to emulate the channel gain of backscatter links.
Here, we assume the backscatter performance is monotonic~\cite{inNout-mobicom20, remix-sigcomm18}, i.e., the backscatter output power is proportional to the received (input) power.
Note that the control algorithm runs with the signal strength of the backscatter channel as feedback instead of separate channels.
We place the transmitter and receivers at 3 locations and test 15 in-water locations for backscatter, measuring 45 backscatter links in total.
As shown in \figref{fig:eval-backscatter-gain}, our system provides a median gain of 10.3~dB and up to 30~dB for received backscatter signal power.
The gain comes from both transmitter-to-backscatter (air-to-water) and backscatter-to-receiver (water-to-air) channels.

\begin{figure*}[t]
  \begin{minipage}{.32\textwidth}
    \centering
    \includegraphics[width=0.95\columnwidth]{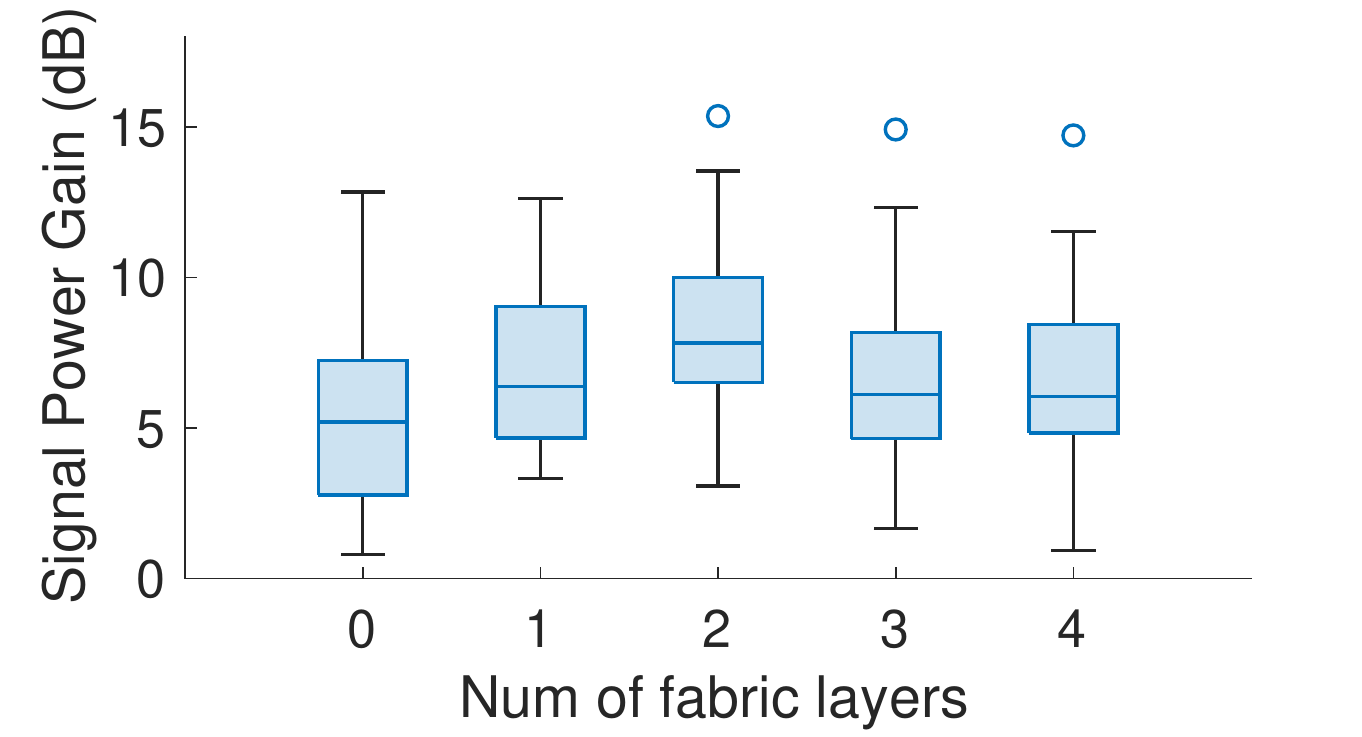}
    \caption{Performance with different surface-media gap distances.}
    \label{fig:eval-water-gain-vs-gap}
  \end{minipage}\hfill
  \begin{minipage}{.32\textwidth}
    \centering
    \includegraphics[width=0.95\columnwidth]{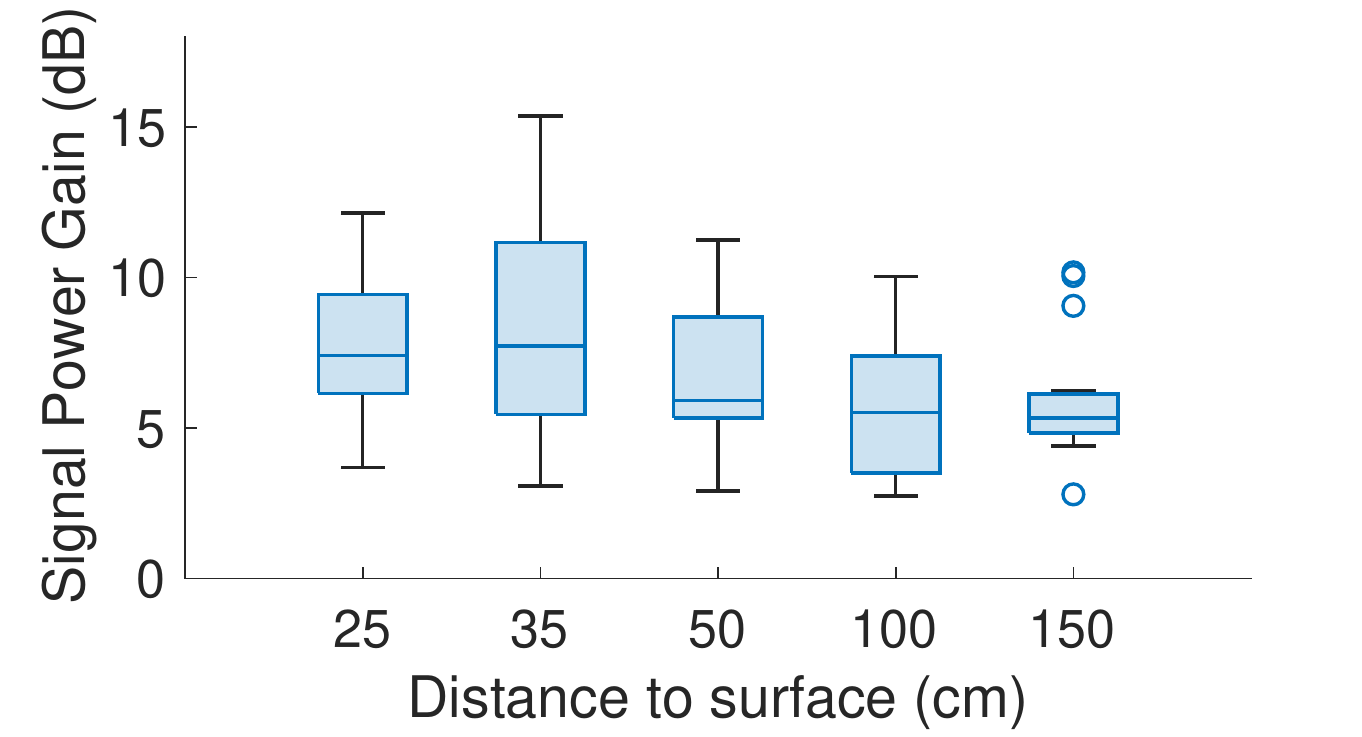}
    \caption{Performance with different surface-endpoint distance.}
    \label{fig:eval-water-gain-vs-distance}
  \end{minipage}\hfill
  \begin{minipage}{.32\textwidth}
    \centering
    \includegraphics[width=0.95\columnwidth]{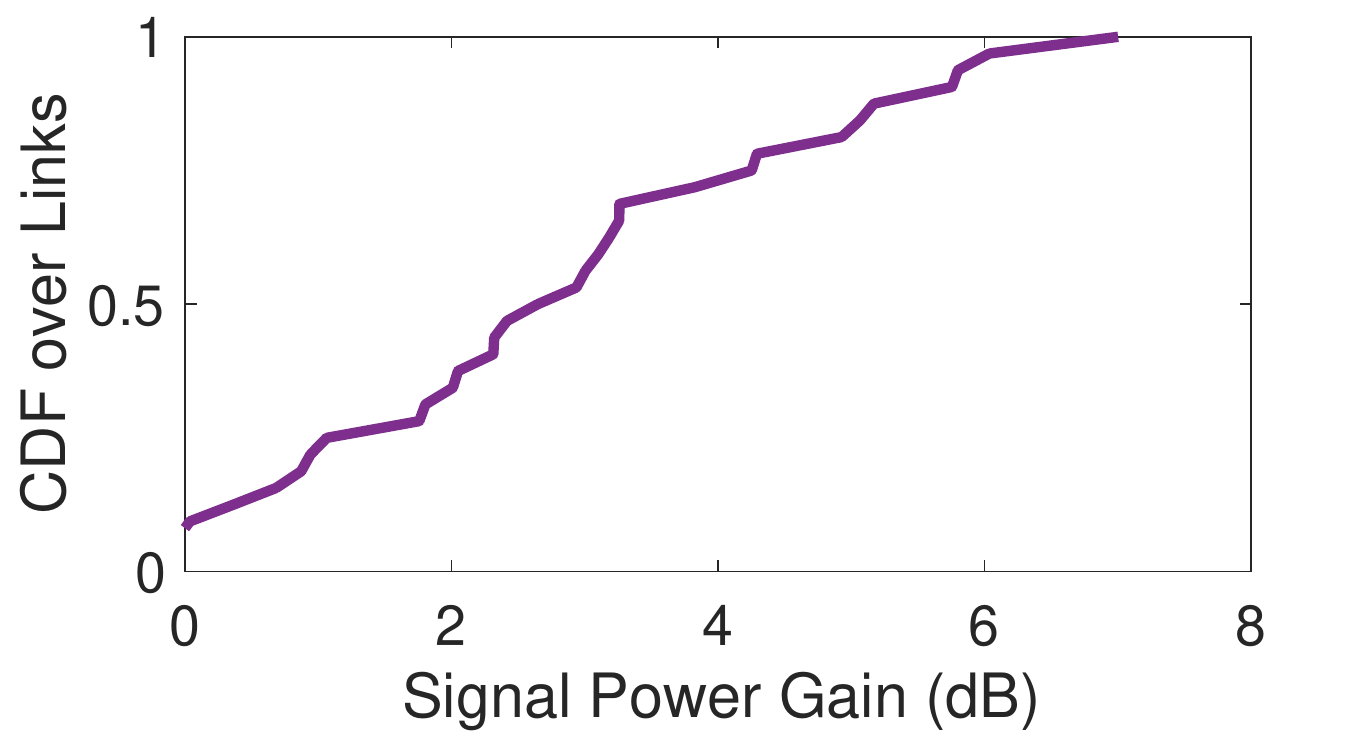}
    \caption{Signal power gain for in-air links.}
    \label{fig:eval-reflex-air-gain}
  \end{minipage}
\end{figure*}

\heading{Surface-medium gap.}
The gap between the surface and the medium can hardly stay fixed at the best distance for most application scenarios due to user movements or other practical deployment constraints.
Therefore, we next evaluate the system performance when adapting to changing surface-water gap sizes.
To accurately control the gap between the flexible surface and media, we change the gap distance with different numbers of fabric layers between the surface and water bucket.
Each fabric layer adds around 2~mm and the bucket has around 2~mm thickness.
We test fabric layers from 0 to 4, i.e., 2 to 10~mm gap, with 15 air-to-water links for each setup.
The results in \figref{fig:eval-water-gain-vs-gap} show over 5~dB median gain and up to 15~dB gain for all gap distances, which demonstrates the effectiveness of surface adaptation.
We note that performance is best with 2 fabric layers (6~mm) and worst with 0 fabric layer (2~mm). 
This matches well with simulation results (\figref{fig:design-hfss-water-gap-heatmap}) that the realizable matching performance varies with different gap sizes, although a high gain can be achieved regardless of gap size.

\heading{Distance to endpoint.}
As the surface can not cover the whole media, only part of signals may go through the surface.
We measure 15 air-to-water links with different distances between the surface and the in-air endpoint.
As shown in \figref{fig:eval-water-gain-vs-distance}, the performance of our system degrades slowly but maintains a high gain for signal power as the distance increases.

\subsection{Supporting over-the-air links}
\label{sec:eval-link-air}
Our system can support links operating entirely in the air but close to some media interface as discussed in \secref{sec:design-diss}.
We measure 32 links in the air with one endpoint placed 35~cm in front of the surface and water.
\figref{fig:eval-reflex-air-gain} shows a median gain of 3~dB and up to 7~dB for received signal power.
Compared to existing work supporting in-air links~\cite{laia-nsdi19,rfocus}, 
we achieve a high power gain with a very small piece of surface.

%% file: related.tex
\section{Related Work}
\label{s:related}
\heading{Smart surfaces systems.}
A recent line of research~\cite{laia-nsdi19,press-hotnets17,rfocus,scattermimo,lava-sigcomm21,rflens_2021} aims to control signal propagation behavior with smart surface systems and improve the perceived channel conditions at receivers. 
LAIA uses phase shifters to control signal phase so that signals add constructively at receivers; RFocus takes a simpler design that only performs on-off amplitude control to achieve beamforming effect; LAVA uses active amplify-and-forward elements deployed as a surface on the ceiling. 
However, they are unable to handle cross-media links since it is our of their design consideration.

\heading{Air-tissue networking}
Wireless networking is critical for numerous in-vivo applications, including vital signals monitoring, gastrointestinal diagnostic, drug delievery, neuro-stimulation therapy~\cite{burton_wireless_2020, traverso_physiologic_2015, meron_wireless_2000, lee_biological_2015, sun_closed-loop_2014, schuster_new_2016}.
In-body wireless endpoints are often swallowed as ingestible electronics~\cite{steiger_ingestible_2019} or injected as medical micro-implants~\cite{implant_dev_2013}.

\heading{Air-water networking.}
Tarf~\cite{tarf-sigcomm18} uses the combination of speaker and RF radar to establish a water-to-air link. Amphilight~\cite{amphilight-nsdi2020} uses laser to circulate the difficulty of RF signal propagation. However, both use very application specific approaches that can hardly be used for other cross-media netowrking scenarios. They also require specialized endpoints, which are much less accessible than wireless IoT devices.

\heading{Air-to-ground networking.}
Communication through or close to air-ground interfaces are important for operations such as mining and tunnel rescue~\cite{wireless-in-tunnel, wireless_in_mines, ground_radar_2008}.
Some recent work propose to sense soil properties like moisture with RF signals~\cite{ding_soil_sensing,khan_estimating_soil_2022}.
Our system can be a enhancement for these applications.

%% file: conclusion.tex

\section{Conclusion} 
\label{s:concl}

As more wireless IoT applications involve cross-media links, they face the fundamental challenge of signal degradation when waves propagate through media interfaces. In this paper, we present a programmable metasurface system, \name, to tackle the root cause of the issue and dynamically match the signal propagation characteristics of the media on both sides of the interface, as if no physical interface existed. With a simple hardware design minicking an array of individually controlled patch antennas, fabricated on one layer of thin, flexible plastic, our prototype is suitable for a range of practical applications. We believe this provides an enabling technology for future IoT applications in challenging communication settings.